\documentclass[aip,amsmath,amssymv,reprint]{revtex4-2}
\usepackage{graphicx}
\usepackage{float}
\usepackage{subfig}
\graphicspath{{./figures/}}

\usepackage{dcolumn}
\usepackage{bm}

\usepackage[utf8]{inputenc}
\usepackage[T1]{fontenc}
\usepackage{mathptmx}
\usepackage{etoolbox}

\usepackage[version=4]{mhchem}
\usepackage{braket}
\usepackage{multirow} 

\usepackage{subfiles}
\usepackage{todonotes}
\usepackage{amsmath}
\usepackage{ulem}
\usepackage{orcidlink}

\usepackage{xr}
\newcommand{\Endo}[2]{\ce{#1}@\ce{#2}}
\newcommand{\wavenumber}{cm$^{-1}$ }

\makeatletter
\def\@email#1#2{%
 \endgroup
 \patchcmd{\titleblock@produce}
  {\frontmatter@RRAPformat}
  {\frontmatter@RRAPformat{\produce@RRAP{*#1\href{mailto:#2}{#2}}}\frontmatter@RRAPformat}
  {}{}
}%
\makeatother

\begin{document}
    \title[\Endo{He}{C60} 4D]{Translational eigenstates of \Endo{He}{C60} from four-dimensional \textit{ab initio} Potential Energy Surfaces interpolated using Gaussian Process Regression} 
    \author{K. Panchagnula*\orcidlink{0009-0004-3952-073X}}
    \email{ksp31@cam.ac.uk}
     \affiliation{Yusuf Hamied Department of Chemistry, University of Cambridge, Cambridge, United Kingdom}
    \author{D. Graf \orcidlink{0000-0002-7640-4162}}
    \author{F.E.A. Albertani}
    \author{A.J.W. Thom \orcidlink{0000-0002-2417-7869}}%

    \date{\today}

    \begin{abstract}
        We investigate the endofullerene system \Endo{^3He}{C60} with a four-dimensional Potential Energy Surface (PES) to include the three  \ce{He} translational degrees of freedom and \ce{C60} cage radius. We compare MP2, SCS-MP2, SOS-MP2,  RPA@PBE and C(HF)-RPA to calibrate and gain confidence in the choice of electronic structure method. Due to the high cost of these calculations, the PES is interpolated using Gaussian Process Regression (GPR), owing to its effectiveness with sparse training data. The PES is split into a two-dimensional radial surface, to which corrections are applied to achieve an overall four-dimensional surface. The nuclear Hamiltonian is diagonalised to generate the in-cage translational/vibrational eigenstates. The degeneracy of the three-dimensional harmonic oscillator energies with principal quantum number $n$ is lifted due to the anharmonicity in the radial potential. The $(2l+1)$-fold degeneracy of the angular momentum states is also weakly lifted, due to the angular dependence in the potential. We calculate the fundamental frequency to range between 96\wavenumber and 110\wavenumber\, depending on the electronic structure method used. Error bars of the eigenstate energies were calculated from the GPR and are on the order of approximately $\pm$ 1.5\wavenumber. Wavefunctions are also compared by considering their overlap and Hellinger distance to the one-dimensional empirical potential. As with the energies, the two \textit{ab initio} methods MP2 and RPA@PBE show the best agreement. While MP2 has better agreement than RPA@PBE, due to its higher computational efficiency and comparable performance, we recommend RPA as an alternative electronic structure method of choice to MP2 for these systems.

    \end{abstract}

    \maketitle
    
    \section{Introduction \label{sec:Intro}}    
    	 Endofullerenes are a class of systems where atom(s) or small molecule(s) are trapped inside fullerene cages. While the development of the technique known as ``molecular surgery''\cite{murataSynthesisReactionFullerene2008, bloodworthSynthesisEndohedralFullerenes2022} has allowed for synthesis and characterisation of these species experimentally they are also of interest from a theoretical perspective\cite{bacicPerspectiveAccurateTreatment2018}. The encapsulating cage provides a confining potential experienced by the endohedral species, which in turn quantises its translational motion. 

    The simplest examples of these systems that do not have any other degrees of freedom, are monoatomic noble gas endofullerenes. Among these, \Endo{He}{C60} has been the subject of recent experimental and theoretical investigation.\cite{bacanuExperimentalDeterminationInteraction2021c, jafariTerahertzSpectroscopyHelium2022} Bacanu et al. synthesised both \Endo{^3He}{C60} and \Endo{^4He}{C60}, and characterised them using using both Terahertz (THz) spectroscopy and Inelastic Neutron Scattering (INS). From these spectra, they simplified the model of the potential energy surface (PES) by considering it to be spherically symmetric, and neglecting the influence of the endohedral \ce{He} atom on the \ce{C60} cage radius. This reduces the complexity of the PES to a single dimension, specifically the distance of the \ce{He} atom from the centre of the cage.\cite{bacanuExperimentalDeterminationInteraction2021c, jafariTerahertzSpectroscopyHelium2022}

    This work extends upon this treatment by incorporating the angular dependence and allows the cage radius to vary. A consequence of this is that the PES becomes four-dimensional: three translational degrees of freedom of the encapsulated \ce{He} atom, and the \ce{C60} isotropic cage breathing mode.

    The predominantly dispersive interaction between the \ce{He} atom and \ce{C60} cage\cite{bacicPerspectiveAccurateTreatment2018}, combined with the high dimensionality of this PES imposes substantial demands on the electronic structure routines. It specifically requires the use of methods that are both very accurate in accounting for dispersion effects, and computationally efficient.

    To reduce the computational workload, frequently a functional form of the surface is chosen, and expansion coefficients are fit to the generated data.\cite{xuH2HDD22008, xuQuantumDynamicsCoupled2008, xuCoupledTranslationrotationEigenstates2009, xuInelasticNeutronScattering2013, felkerCommunicationQuantumSixdimensional2016, kaluginaPotentialEnergyDipole2017, xuLightMoleculesNanocavities2020, felkerFlexibleWaterMolecule2020, jafariNeArKr2023a, panchagnulaExploringParameterSpace2023b} However, this procedure is very rigid and can depend strongly on the functional form chosen and its parameters.

    Recent years have seen major developments in approaches to the interpolation problem based on machine learning frameworks. These allow for the prediction of PESs without the necessity of strong \textit{a priori} knowledge of all its features. For example, Gaussian Processes (GPs) can span accurate surfaces for very different systems, despite using a very common, generic construction.\cite{noeMachineLearningMolecular2020, deringerGaussianProcessRegression2021} This is possible due to the fact that GPs adapt their underlying characteristics to those of the true surface. The amount of data required for accurate interpolations depends on the dimensionality of the PES, the hypervolume needed to cover the relevant region of the surface and the roughness of the surface. Smooth and regular surfaces, which PESs are, can be accurately described by quite small, sparse datasets.\cite{loeppkyChoosingSampleSize2009}
    
    It is no surprise that GPs have been extensively used in computational chemistry to generate cheaper PES evaluations for both small systems,\cite{behlerPerspectiveMachineLearning2016, musilPhysicsInspiredStructuralRepresentations2021, dralStructurebasedSamplingSelfcorrecting2017} larger scale chemical simulations,\cite{westermayrPerspectiveIntegratingMachine2021} excited states\cite{westermayrMachineLearningElectronically2021} and reaction dynamics.\cite{collinsMolecularPotentialenergySurfaces2002a} The biggest challenge in creating accurate descriptions of these surfaces is not to explicitly obtain the data but to make it efficiently predict new data. Instead, one is usually hit with the curse of dimensionality due to the exponential increase in the hypervolume with increasing input dimensions. Therefore using efficient descriptors for the input data is crucial.\cite{bartokMachinelearningApproachOne2013, behlerPerspectiveMachineLearning2016, albertaniGlobalDescriptorsWater2023, albertaniOptimisedMorseTransform2023}

    Once armed with the PES, the energies of the translational states can be found by diagonalising the nuclear Hamiltonian. While this gives the full ``absolute'' energy spectrum, only the differences between energy levels are of physical importance, as these are the transitions observed spectroscopically. The comparison of these energy gaps with the previously reported one-dimensional and spectroscopic data\cite{bacanuExperimentalDeterminationInteraction2021c, jafariTerahertzSpectroscopyHelium2022} distinguishes the quality of the PES, and by consequence the quality of its underlying \textit{ab initio} method.

    With a spherically symmetric potential, using a basis set of three-dimensional harmonic oscillator functions generates energies which are $(2l+1)$-fold degenerate. The angular dependence in the potential allows coupling of states with differing angular momentum quantum numbers and this effect is observed in the lifting of degeneracy of these states as expected under $I_h$ symmetry. The importance of the angular dependence in the potential can also be ascertained by analysing the wavefunctions, and their deviations from the purely spherically symmetric states. The influence of the cage breathing mode can be determined analogously.

    The rest of this work is structured as follows: Section \ref{sec:Methodology}
    contains the theory and methodology used to calculate the translational eigenstates of \Endo{^3He}{C60}. This is split into Section \ref{sec: Theory ES} which provides an explanation of the rationale of the choices of electronic structure methods; Section \ref{sec: Theory GP} details the construction of the four-dimensional PES using GPs expanded in spherical polar coordinates; and Section \ref{sec: Theory EF} outlines the calculation of the translational eigenstates of the system. The results of this methodology and a discussion are presented in Section \ref{sec:Results}. Concluding remarks and prospective avenues for future research are outlined in Section \ref{sec:Conc}.

    \section{Theory \label{sec:Methodology}}
    	\subsection{Electronic Structure Calculations\label{sec: Theory ES}}
        \subsubsection{Choosing an appropriate method}
    
            In \Endo{He}{C60} the choice of electronic structure method is ultimately a balancing act between accurately depicting the effects of dispersion, and computational efficiency. A lot of previous research on endofullerenes forgoes this, by employing a pairwise-additive LJ potential, summed over the cage-endohedral atom interactions, with the parameters derived empirically. \cite{xuH2HDD22008, xuQuantumDynamicsCoupled2008, xuCoupledTranslationrotationEigenstates2009, xuInelasticNeutronScattering2013, felkerCommunicationQuantumSixdimensional2016, xuLightMoleculesNanocavities2020, felkerFlexibleWaterMolecule2020, jafariNeArKr2023a, panchagnulaExploringParameterSpace2023b} However, the choice of these parameters may not be unique, and their optimal values are strongly dependent on the environmental conditions.\cite{xuCoupledTranslationrotationEigenstates2009, panchagnulaExploringParameterSpace2023b} Recent work also suggests these empirical models yield potentials with considerable variance and can often lead to poor agreement with experimental observations. This is not a surprise due to the delocalised electronic structure of the \ce{C60} cage. \cite{bacanuExperimentalDeterminationInteraction2021c, jafariTerahertzSpectroscopyHelium2022, jafariNeArKr2023a}

            Density functional theory (DFT) emerges as a natural alternative to this procedure, due to its renowned success across computational chemistry, physics and materials science. This is primarily owing to its favourable cost-to-performance ratio, which is why it is commonly referred to as the workhorse of quantum chemistry.\cite{tealeDFTExchangeSharing2022}
            Nonetheless, standard density functional approximations (DFAs) have well-documented limitations in accurately capturing dispersion forces.\cite{perez-jordaDensityfunctionalStudyVan1995} To mitigate this, various techniques including Grimme's dispersion corrections \cite{grimmeDispersionCorrectedMeanFieldElectronic2016, grimmeAccurateDescriptionVan2004, antonyDensityFunctionalTheory2006, grimmeSemiempiricalGGAtypeDensity2006, grimmeConsistentAccurateInitio2010, caldeweyherExtensionD3Dispersion2017, caldeweyherGenerallyApplicableAtomiccharge2019} have been developed and have gained enormous popularity. However, given the extensive array of DFAs,\cite{mardirossianThirtyYearsDensity2017} some of them yielding divergent results in prior endofullerene studies,\cite{kaluginaPotentialEnergyDipole2017} the reliability of these methods in accurately modelling these systems remains questionable.

            On the other hand, wavefunction (WF) methods are reputed for their ability to deliver highly accurate results, albeit with an unforgiving relationship between accuracy and computational expense. The significant computational demand arises from both the unfavourable scaling with system size, and also the inherent slow convergence with respect to basis set size.\cite{ginerCuringBasissetConvergence2018} The latter effect is due to the necessity that WF methods explicitly describe the electronic cusp\cite{katoEigenfunctionsManyparticleSystems1957} which requires the use of many basis functions with a large angular momentum quantum number.\cite{fabianoAccuracyBasissetExtrapolation2012} DFAs on the other hand, implicitly incorporate this condition through exchange-correlation functionals leading to faster basis set convergence.

            Nevertheless, significant advancements have been made in the field of WF electronic structure methods over recent decades. A prime example of this is second order M\o{}ller--Plesset perturbation theory (MP2) which stands out for its efficiency. \cite{pulayLocalizabilityDynamicElectron1983, haserMollerPlessetMP2Perturbation1993, maslenNoniterativeLocalSecond1998,ayalaLinearScalingSecondorder1999,schutzLoworderScalingLocal1999, saeboLowscalingMethodSecond2001, wernerFastLinearScaling2003, jungFastCorrelatedElectronic2006,jungFastEvaluationScaled2007, doserTighterMultipolebasedIntegral2008, doserLinearscalingAtomicOrbitalbased2009, zienauCholeskydecomposedDensitiesLaplacebased2009, kristensenMP2EnergyDensity2012, maurerCholeskydecomposedDensityMP22014, pinskiSparseMapsSystematic2015,nagyIntegralDirectLinearScalingSecondOrder2016, baudinEfficientLinearscalingSecondorder2016, phamHybridDistributedShared2019, barcaQMP2OSMollerPlesset2020, glasbrennerEfficientReducedScalingSecondOrder2020, forsterQuadraticPairAtomic2020} Notably, the only current higher dimensional PES of an endofullerene, \Endo{HF}{C60}, was constructed using density-fitting local MP2\cite{kaluginaPotentialEnergyDipole2017} developed by Werner et al.\cite{wernerFastLinearScaling2003} In our work, we will utilise an advanced MP2 variant known as  $\omega$-RI-CDD-MP2, a method recently introduced by the Ochsenfeld group.\cite{glasbrennerEfficientReducedScalingSecondOrder2020} The specific thresholds applied are outlined within the supplementary information in Section SI 1.

            We note, with caution, that there is a growing body of evidence indicating that MP2 can substantially overestimate non-covalent interactions, with discrepancies reaching up to 100\%. \cite{hohensteinWavefunctionMethodsNoncovalent2012, janowskiAccurateCorrelatedCalculation2010, janowskiBenchmarkQuantumChemical2011, grimmeConsistentAccurateInitio2010, grimmeSupramolecularBindingThermodynamics2012} These inaccuracies tend to systematically increase with system size, which suggests a strong concern in applying these methods to large dispersion-bound complexes.\cite{nguyenDivergenceManyBodyPerturbation2020a} The root of MP2's shortcomings is the inadequate treatment of electrodynamic screening of the Coloumb interactions among electrons induced by particle-hole pairs resulting in an overestimation of the effective interaction strength.\cite{nguyenDivergenceManyBodyPerturbation2020a}
        
            Concerned by these findings, we decided against relying exclusively on the MP2 method. Research by Furche and colleagues has demonstrated that while the scaled opposite spin (SOS) and spin component scaled (SCS) MP2 methods have similar issues, their errors are generally less severe.\cite{nguyenDivergenceManyBodyPerturbation2020a} Consequently, we will also include results from SOS-MP2 and SCS-MP2 in our discussion.

            An additional method that does account for correlation by including electrodynamic screening through induced particle-hole pairs or density fluctuations, thereby reducing the effective electron interaction, is the random phase approximation 
            (RPA).\cite{nguyenDivergenceManyBodyPerturbation2020a} RPA can be interpreted as a resummation of all possible ring diagrams, and due to the neglect of exchange contributions between particle-hole pairs it can be evaluated even more efficiently than the MP2 method. The RPA has been recognised for counteracting the erratic behaviour of MP2, proving to be reliable across various system sizes.\cite{nguyenDivergenceManyBodyPerturbation2020a} With its good description of long-range correlation effects such as dispersion, RPA therefore serves as an excellent tool for validating our results. In the present work, we will hence utilise the linear scaling $\omega$-CDGD-RPA method put forward by the Ochsenfeld group which has been shown to provide excellent agreement with conventional RPA.\cite{grafAccurateEfficientParallel2018}
            While self-consistent formulations of RPA have been introduced in the literature,\cite{vooraVariationalGeneralizedKohnSham2019, yuSelfconsistentRandomPhase2021, grafLowScalingSelfConsistentMinimization2019, grafRangeseparatedGeneralizedKohn2020} RPA is predominantly applied in a post-Kohn--Sham manner,\cite{furcheMolecularTestsRandom2001} utilising orbitals and orbital energies obtained from a preceding DFA calculation, which we will abbreviate as RPA@DFA in the following. Typically, these reference calculations employ a generalized gradient approximation (GGA), with the one proposed by Perdew, Burke, and Ernzerhof (PBE)\cite{perdewGeneralizedGradientApproximation1996, perdewGeneralizedGradientApproximation1997} being particularly popular. However, pure density functionals such as GGAs are known for their self-interaction error \cite{cohenInsightsCurrentLimitations2008, simImprovingResultsImproving2022} which, in turn, can lead to erroneous densities, Kohn--Sham orbitals, and orbital energies, impacting the subsequent RPA calculation. Aware of this issue, we also present results obtained with the corrected Hartree--Fock RPA (C(HF)-RPA) approach, designed to address these errors.\cite{grafCorrectedDensityFunctional2023}

        \subsubsection{Ensuring Basis-Set Convergence\label{sec: Theory CBS}}
    
            WF methods generally exhibit slow convergence with respect to the basis-set size. The RPA, sitting on the border between WF theory and DFT, shows similar limitations; in fact, research suggests that it converges even slower than MP2.\cite{eshuisBasisSetConvergence2012, fabianoAccuracyBasissetExtrapolation2012} Given the pronounced basis set incompleteness error in dispersion-bound systems,\cite{eshuisBasisSetConvergence2012} extrapolation to the complete basis set (CBS) limit is strongly advised for both MP2 and RPA.
        
            It was shown that the counterpoise correction\cite{boysCalculationSmallMolecular1970} does not
            lead to significant improvements and further that augmenting the basis set can even slow down the basis set convergence without offering substantial benefits.\cite{eshuisBasisSetConvergence2012, fabianoAccuracyBasissetExtrapolation2012} We therefore use the popular two-point extrapolation\cite{eshuisBasisSetConvergence2012,fellerEffectivenessCCSDComplete2011,halkierBasissetConvergenceCorrelated1998,helgakerBasissetConvergenceCorrelated1997}

                \begin{equation}
                    E^{\infty}_{\text{corr}} = \frac{E_n (n + d)^3 - E_m (m + d)^3}
                    {(n + d)^3 - (m + d)^3} \label{eq: CBS correlation}
                \end{equation}
            with $n = 5$, $m = 4$, and $d = -1.17$ or $d = 0.0$ for RPA and MP2 respectively, in combination with Dunning's correlation-consistent polarised quadruple-$\zeta$ (cc-pVQZ) and quintuple-$\zeta$ (cc-pV5Z) valence basis sets.\cite{balabanovSystematicallyConvergentBasis2005, dunningGaussianBasisSets1989, koputInitioPotentialEnergy2002, prascherGaussianBasisSets2011, wilsonGaussianBasisSets1999, woonGaussianBasisSets1993, woonGaussianBasisSets1994}

            For the SCF energies, the extrapolations were carried out using the three-point formula advocated by Feller\cite{fellerUseSystematicSequences1993}

                \begin{equation}
                    E_{\text{SCF}}^{X} = E_{\text{SCF}}^{\infty} + a \exp\left(-bX\right) \label{eq: CBS SCF}
                \end{equation}

            using the correlation consistent basis sets cc-pVXZ\cite{balabanovSystematicallyConvergentBasis2005, dunningGaussianBasisSets1989, koputInitioPotentialEnergy2002, prascherGaussianBasisSets2011, wilsonGaussianBasisSets1999, woonGaussianBasisSets1993, woonGaussianBasisSets1994} with X$\in\{\text{T, Q, 5}\}$. For additional computational details, readers are directed to the supplementary information, Section SI 1.

    \subsection{Generating the Potential Energy Surface\label{sec: Theory GP}}
        \subsubsection{Gaussian Process Regression}

            A Gaussian Process (GP) is a machine learning regression method, and is defined as \textit{a collection of random variables, any finite number of which have a joint Gaussian distribution}.\cite{rasmussenGaussianProcessesMachine2005} An essential part of a GP model is its covariance function, or kernel, which is equivalent to a similarity measure of the data scattered throughout an input space.

            Kernels may take a wide variety of mathematical forms, as they only need to adhere to a few simple rules.\cite{rasmussenGaussianProcessesMachine2005} More complex kernels can be constructed by multiplying and/or adding other kernels and the rule
                \begin{equation}
                    \mathrm{K}(\mathbf{X}, \mathbf{X}') = \mathrm{K}_{1a}(\mathbf{X}, \mathbf{X}')\mathrm{K}_{1b}(\mathbf{X}, \mathbf{X}')+\mathrm{K}_2(\mathbf{X}, \mathbf{X}')\label{eq: Kernel}
                \end{equation}
            yields a valid ``composite'' kernel function.

            A GP is trained on a set of training data, with inputs $\mathbf{X}_t$ with known outputs $y(\mathbf{X}_t)$ and predicts and multivariate Gaussian distribution defined by the mean $\mu(\mathbf{X}_p)$ and covariance $\Sigma(\mathbf{X}_p)$\cite{rasmussenGaussianProcessesMachine2005}
                \begin{align}
                    \mu(\mathbf{X}_p) &= \mathbf{K}_{pt}\mathbf{K}_{tt}^{-1}\mathbf{y}_t \label{eq: GP mean}\\
                    \Sigma(\mathbf{X}_p) &= \mathbf{K}_{pp} -\mathbf{K}_{pt}\mathbf{K}_{tt}^{-1}\mathbf{K}_{tp} \label{eq: GP covariance} 
                \end{align}
            where $\mathbf{K}_{ij}$ is shorthand for the kernel $K(\mathbf{X}_i, \mathbf{X}_j)$ and the subscripts $t$ and $p$ are for the training and prediction points respectively. A common metric used by the machine learning community to gauge the efficacy of the GP is to consider the 95\% confidence interval of the prediction, which is given by $\mu(\mathbf{X}_p) \pm 1.96\Sigma(\mathbf{X}_p)$.

            GPs are trained by finding an optimal set of hyperparameters, so-called as to differentiate them between parameters in the kernel which are left unchanged during the optimisation process. These hyperparameters are what introduce the flexibility into the model, and can control length scales, amplitude and noise within the data. Using a Bayesian approach, this hyperparamter set is found by maximising the log-marginal likelihood (LML) given by\cite{rasmussenGaussianProcessesMachine2005} 
                \begin{equation}
                    \mathrm{LML} = -\frac{1}{2} \mathbf{y}_t^{\mathrm{T}} \mathbf{K}_{tt}^{-1}\mathbf{y}_t - \frac{1}{2} \log|\mathbf{K}_{tt}| - \frac{n}{2} \log (2\pi).
                    \label{eq:lml}
                \end{equation}
            The three contributing terms to this quantity are to be understood as a fit, a regularisation, and a normalisation factor.

        \subsubsection{Building an appropriate interpolation}

\begin{figure}
                \centering
                \includegraphics[scale=0.2]{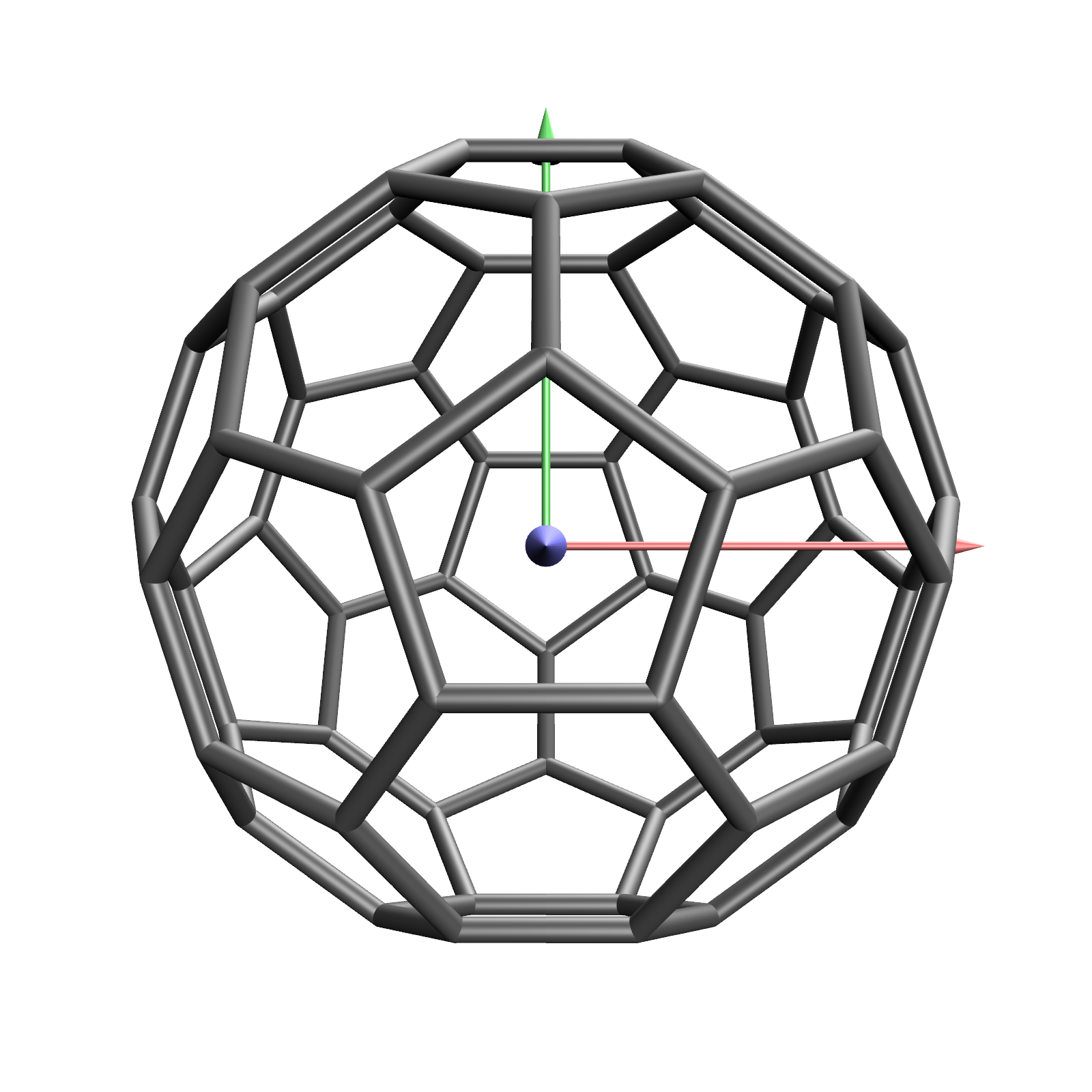}
                \caption{Axis orientation of \ce{C60}. Red, green and blue arrows correspond to $x$, $y$ and $z$ directions respectively}
                \label{fig: C60}
            \end{figure}

            The high $I_h$ symmetry of the \Endo{He}{C60} system, motivates the use of a spherical polar coordinate system ($r_{\ce{He}}$, $\theta$, $\phi$, $R_{\ce{C60}}$), with $(\theta, \phi)$ representing the polar and azimuthal angles respectively. The \ce{C60} cage is oriented as given in Fig \ref{fig: C60}, with the $z$-axis going through the centre of a pentagon, a $C_5$ rotation axis, and the $y$-axis aligned through the centre of a hexagon-pentagon \ce{C-C} bond.

            Analysis of the spectroscopic data\cite{bacanuExperimentalDeterminationInteraction2021c, jafariTerahertzSpectroscopyHelium2022} indicates that the PES is dominated by the distance of the encapsulated \ce{He} atom from the centre of the cage. This motivates the partitioning of this PES as a two-dimensional ($r_{\ce{He}}$, $R_{\ce{C60}}$) radial surface to which four-dimensional ($r_{\ce{He}}$, $\theta$, $\phi$, $R_{\ce{C60}}$) corrections are added. These corrections are on the order 1\wavenumber, and add some fluctuations to an otherwise smooth, polynomial radial surface. 

            This form of PES necessitates the learning of two separate but linked GPs. Initially, the two-dimensional radial surface is trained on the majority of the data. The four-dimensional corrections are then trained the remaining input data, but with $y_{4D}=y_{2D}-\mu_{2D}$. That is to say, its outputs are the difference between the learned radial surface and electronic structure data. Finally, before training the four-dimensional surface, the input coordinates are symmetrised. This is accomplished by applying the $I_h$ symmetry operations $C_5^n$ for $n\in [0, 1, 2, 3, 4]$ using the axis aligned along the $z$ direction, and also the centre of inversion $i$, producing nine extra equivalent geometries with identical energies.

            The quality of the PES is determined by the choice of kernel and distance metric used in the GP. For the two-dimensional radial surface, an anisotropic Mat\'{e}rn covariance function\cite{rasmussenGaussianProcessesMachine2005} with $\nu=2.5$ using the standard Euclidean distance added to a white noise function was chosen. For the four-dimensional surface, a more considered choice is required as the angular and radial components use different types of coordinates, units and possibly even distance metrics. Consequently, this kernel is split into the sum of a white noise function and a product of a radial and angular functions
                \begin{align}
                    K_{2D}(\mathbf{r},\mathbf{r}') &= \sigma^2_{2D}k(\mathbf{r},\mathbf{r}') + \text{Noise} \label{eq: 2D kernel}\\
                    K_{4D}(\mathbf{R},\mathbf{R}') &= \sigma^2_{4D}k_{r}(\mathbf{r},\mathbf{r}')k_{a}(\boldsymbol{\Omega},\boldsymbol{\Omega}') +\text{Noise}. \label{eq: 4D kernel}
                \end{align}
            where $\mathbf{r}$ refers to the radial coordinates ($r_{\ce{He}}$, $R_{\ce{C60}}$), $\boldsymbol{\Omega}$ refers to the angular coordinates ($\theta$, $\phi$) and $\mathbf{R}:=(\mathbf{r}, \boldsymbol{\Omega})$. The product of $k_r$ and $k_a$ is chosen as it treats both coordinate types on equal footing. An alternative choice of summed kernel here would treat the similarity as similar to the radii or the angles which would introduce artificial correlation with data. 
            
            While $k_r=k$ is a possible choice, a more nuanced approach is required for $k_a$. This is due to the use of spherical polar coordinates, and how this affects the choice of metric. As the angular coordinates (without loss of generality) lie on the surface of a unit sphere this provokes the choice of two metrics: chordal, which is the simple straight line distance; or great-circle, which is the geodesic on the sphere. This is calculated as 

                \begin{align}
                    & d_{\text{Great-Circle}}(\boldsymbol{\Omega}, \boldsymbol{\Omega'}) = 2\arcsin\left(\left[\sin^2(\frac{(\frac{\pi}{2}-\theta)-(\frac{\pi}{2}-\theta')}{2})\right. \right. \nonumber \\
                     &+\left. \left.\cos(\frac{\pi}{2}-\theta)\cos(\frac{\pi}{2}-\theta')\sin^2(\frac{\phi-\phi'}{2})\right]^{0.5}\right) \label{eq: Great Circle Distance}\\
                     & d_{\text{Chordal}} = 2 \sin\left(\frac{d_{\text{Great-Circle}}}{2}\right) \label{eq: Chordal-GreatCircle}
                \end{align}
            where the spherical polar coordinates have been converted to longitude and latitude. This is easily converted to the chordal distance using the relationship given in Eq \eqref{eq: Chordal-GreatCircle}. If using the chordal metric, as this is analogous to the Euclidean distance of Cartesian coordinates, the usual choices of kernels are possible. However, should the great-circle distance be preferred, this is not the case as a positive-definite covariance matrix is not guaranteed. This is resolved by using compactly supported covariance functions\cite{gneitingStrictlyNonstrictlyPositive2013, padonouPolarGaussianProcesses2016}, examples of which are given in Table \ref{table: Spherical kernels}. These functions can also use the chordal distance, but as the geodesic does not distort these distances, that should be the preferred metric where possible. Due to the high symmetry of the endofullerene, this also motivates the use of a kernel that can emulate the radial and angular patterns within the input data, which is better achieved through the kernels listed in Table \ref{table: Spherical kernels}. 

            \begin{table}
                \centering
                \begin{tabular}{ccc}
                     \hline Kernel& Function & Geodesic Parameters  \\
                     \hline Mat\'{e}rn&$\frac{2^{1-\nu}}{\Gamma(\nu)}\left(\sqrt{2\nu}\frac{d}{l}\right)^{\nu}K_{\nu}\left(\sqrt{2\nu}\frac{d}{l}\right)$&$l>0;\,\nu\in(0,\frac{1}{2}]$\\
                     Wendland-C2&$\left(1+\tau\frac{d}{s}\right)\left(1-\frac{d}{s}\right)^{\tau}_{+}$&$s\in(0,\pi],\,\tau\geq4$\\
                     Wendland-C4&$\left(1+\tau\frac{d}{s}+\frac{\tau^2-1}{3}\frac{d^2}{s^2}\right)\left(1-\frac{d}{s}\right)^{\tau}_{+}$&$s\in(0,\pi],\,\tau\geq6$\\
                     \hline
                \end{tabular}
                \caption{Kernels which give a positive-definite covariance matrix when using the spherical geodesic distance for $d$. In the Mat\'{e}rn, $\nu, l$ refer to the smoothness and length-scale. In the Wendland kernels, $\tau, s$ play analogous roles although $s$ is a support parameter and may not be optimisable. The subscript + is shorthand for max(f(x),0). If using the chordal metric, some parameter restrictions are lifted.}
                \label{table: Spherical kernels}
            \end{table}

            As we require the PES to be twice-differentiable, given the restriction on $\nu$ in Table \ref{table: Spherical kernels} this removes the Mat\'{e}rn as a possibility. The Wendland kernels are dependent on two parameters: $\tau$, a smoothness parameter which is analogous to $\nu$ in the Mat\'{e}rn; and $s$ which appears to be an optimisable lengthscale but is actually a support parameter. To ensure differentiability of the Wendland kernel, we require $s\geq\text{max}(d)$ which is $\pi$ and 2, for the great-circle and chordal distances respectively. While in the latter case this could be an optimised hyperparameter, in both cases we choose to fix $s=\text{max}(d)$.
            
    \subsection{Diagonalising the Translational Hamiltonian\label{sec: Theory EF}}
        \subsubsection{Full 4D Problem}
            The Hamiltonian for \Endo{He}{C60} including all degrees of freedom of the endohedral atom and the cage breathing in atomic units is given by

                \begin{align}
                	  \hat{H} &= \underbrace{-\frac{1}{2m}\nabla^2_{\mathbf{r}_{\ce{He}}}-\frac{1}{2M}\nabla^2_{R_{\ce{C60}}} + \frac{1}{2}k_{\ce{He}}r^2+\frac{1}{2}k_{\ce{C60}} (R_{\ce{C60}}-R_{\text{eq}})^2}_{\hat{H}^0} \nonumber \\
                	  & +\Delta V(\mathbf{R}) \label{eq: 4D Hamiltonian}
                \end{align}
            where $m$ is the reduced mass of the \ce{He}--\ce{C60} ``two-particle'' system calculated as $m=\frac{m_{\ce{He}}m_{\ce{C60}}}{m_{\ce{He}}+m_{\ce{C60}}}$, $M=m_{\ce{C60}}$ is the mass of the \ce{C60} cage, $(r, \theta, \phi)$ refer to the endohedral coordinates of the \ce{He} in spherical polars: distance from origin, polar and azimuthal angles respectively; and $R$ is the \ce{C60} cage radius. The harmonic contributions of the \ce{He} translation and \ce{C60} cage breathing have been explicitly included as this enables the use of the following simple coordinate transformations 
                \begin{align}
                    q_{\ce{He}} & = r^2\sqrt{k_{\ce{He}}m} \label{eq: 3D isotropic He scale}\\
                    q_{\ce{C60}} &= (R_{\ce{C60}}-R_{\text{eq}})\sqrt{\sqrt{k_{\ce{C60}}M}}. \label{eq: 1D isotropic cage scale}
                \end{align}
            where $R_{\text{eq}}$ is the equilibrium \ce{C60} cage radius. By scaling both the \ce{He} and \ce{C60} cage breathing to the natural length scales of the oscillators, and centring them at $q_i=0$, this allows for the basis set to be constructed from a three-dimensional harmonic oscillator (3D HO) for the \ce{He} translation and a one-dimensional harmonic oscillator (1D HO) for the \ce{C60} cage breathing mode. Overall, the basis set is then written as
                \begin{equation}
                    \ket{klmb} = \ket{kl}\ket{lm}\ket{b} = N_{klmb}L_{k}^{l+\frac{1}{2}}(q_{\ce{He}})Y_{lm}(\theta,\phi)H_{b}(q_{\ce{C60}}) \label{eq: 4D basis set}
                \end{equation}
            where $N_{klmb}$ is the appropriate normalisation constant. Working in the finite basis representation (FBR), the $\hat{H}^0$ matrix is diagonal and depends only on the quantum numbers and frequencies of the two different oscillators. The potential energy matrix is a bit more cumbersome to work with. Traditionally this problem is solved using discrete variable representation (DVR) techniques\cite{lightGeneralizedDiscreteVariable1985, lightDiscreteVariableRepresentationsTheir2000, echavePotentialOptimizedDiscrete1992}, which require both the selection of basis set and quadrature points where this matrix would be diagonal, and the quadrature grid would normally be constructed as a direct product of smaller sub-dimensional grids. 

            However, as the angular momentum quantum number $l$ is shared between the radial and angular functions, a direct product grid is not possible. Nonetheless it is still possible to leverage advantages from using the DVR in this scenario. As this basis set is orthonormal, there exists an orthogonal transformation between the FBR and DVR. \cite{lightGeneralizedDiscreteVariable1985} This transformation from the DVR to the FBR corresponds to effectively evaluating the potential matrix elements using Gaussian quadrature.
    
            An appropriate choice of values for $k_{\ce{He}}$ and $k_{\ce{C60}}$ in Equations \eqref{eq: 3D isotropic He scale} and \eqref{eq: 1D isotropic cage scale} will correspond to the equivalent Gaussian quadrature points. There can be multiple choices for this: the traditional method, where an interval for each coordinate is chosen such that the potential doesn't exceed a cutoff values, or a Potential-Optimised DVR\cite{echavePotentialOptimizedDiscrete1992} scheme where the shape and curvature of the potential dictates this scaling. 
    
            For the $q_{\ce{He}}$ coordinate, the atom is restricted to explore a sphere within the cage with $r_{\ce{He}}\in [0,1.5\text{\AA}]$, as even at the equilibrium cage radius beyond this interval the potential is greater than 5000\wavenumber. On the other hand, $q_{\ce{C60}}$ is scaled using the natural known frequency of the cage breathing mode of 496\wavenumber.\cite{bethuneVibrationalRamanInfrared1991}

        \subsubsection{Comparison to 1D}
            After imposing spherical symmetry and fixing the cage radius, the simple 1D translational eigenstates can be calculated from the Hamiltonian\cite{bacanuExperimentalDeterminationInteraction2021c, jafariTerahertzSpectroscopyHelium2022, jafariNeArKr2023a}
                \begin{equation}
                    \hat{H} = -\frac{1}{2m}\nabla^2_{\mathbf{r}_{\ce{He}}} + \frac{1}{2}kr^2 +\Delta V(r). \label{eq: 1D Hamiltonian}
                \end{equation}
            A simple sixth degree polynomial for $V(r)$ has been fitted\cite{bacanuExperimentalDeterminationInteraction2021c, jafariTerahertzSpectroscopyHelium2022}, and by using the $\ket{kl}$ basis set as defined in Equation \eqref{eq: 4D basis set} but with $k$ chosen from a PODVR perspective, the one-dimensional eigenstates can be derived. Anharmonicity lifts the degeneracy of states with the same principal quantum number\cite{bacanuExperimentalDeterminationInteraction2021c, jafariTerahertzSpectroscopyHelium2022} $n=2k+l$, but the lack of angular dependence in this potential implies no coupling of states with differing values of $l$ and maintains their $(2l+1)$-fold degeneracy. Under $I_h$, states with $l\geq3$ are expected to have this degeneracy lifted but this is not observable with the purely one-dimensional radial potential.

            The simplest comparison of the one-dimensional and four-dimensional problems is through the energy gaps between eigenstates as these are what can be measured spectroscopically. The importance of angular dependence in the potential can be determined by how strongly the $(2l+1)$-fold degeneracy of the spherically symmetric states is lifted.

            Not only can the energies of the states be compared, but also the wavefunctions by considering their overlaps
                \begin{align}
                    \braket{\Phi|\Psi}&=\sum_{KLMB}\sum_{klmb}c_{KLMB}c_{klmb}\braket{KLMB|klmb} \nonumber\\                    
                    &=\sum_{KLMB}\sum_{klmb}c_{KLMB}c_{klmb}\braket{KL|kl}\braket{LM|lm}\braket{B|b}\nonumber\\
                    &=\sum_{KLMB}\sum_{klmb}c_{KLMB}c_{klmb}\braket{KL|kl}\delta_{Ll}\delta_{Mm}\braket{B|b} \nonumber \\
                    &=\sum_{KBklmb}c_{KlmB}c_{klmb}\braket{Kl|kl}\braket{B|b}. \label{eq: 4D overlap}
                \end{align}
            While this is the general case for the overlap of two four-dimensional eigenstates, if $\bra{\Phi}$ is the one-dimensional reference, the $\bra{LM}\bra{B}$ functions have to be pre-selected. If they are chosen to be pure basis states, the sum over those indices collapses. The expected orthonormality of the integrals $\braket{Kl|kl}$ and $\braket{B|b}$ is not necessarily recovered due to the different scalings and centres of expansion, and they must be evaluated.

            Instead of using the overlap which measures the similarity of the wavefunctions, an equivalent notion of how far apart the states are can be considered. The Hellinger distance \cite{bhattacharyyaMeasureDivergenceTwo1943}, which is given by
                \begin{equation}
                    H(\Phi,\Psi) = \sqrt{1-|\braket{\Phi|\Psi}|} \label{eq: Hellinger distance}
                \end{equation}
            can be used as an alternative. This requires both wavefunctions to be renormalised along the same measure guaranteeing $0\leq H\leq 1$ with equality holding if the states are completely identical or orthogonal respectively.

    \section{Results \label{sec:Results}}
    	 \subsection{Potential Energy Surface \label{sec: Results PES}}

        \begin{figure*}
            \centering
            \setcounter{subfigure}{0}
            \subfloat[MP2 Radial]{\includegraphics[scale=0.6]{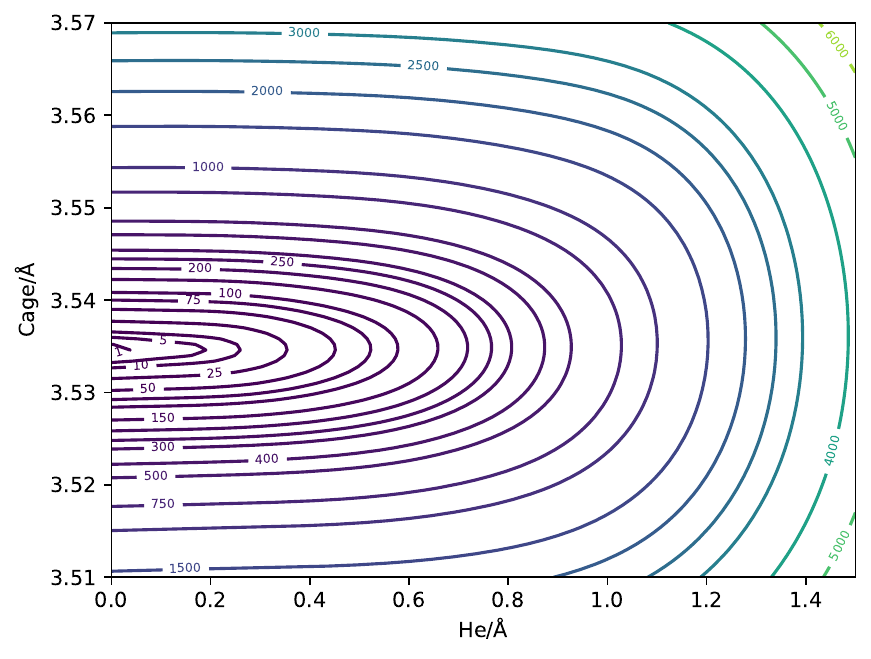}}\,
            \subfloat[MP2 Angular]{\includegraphics[scale=0.6]{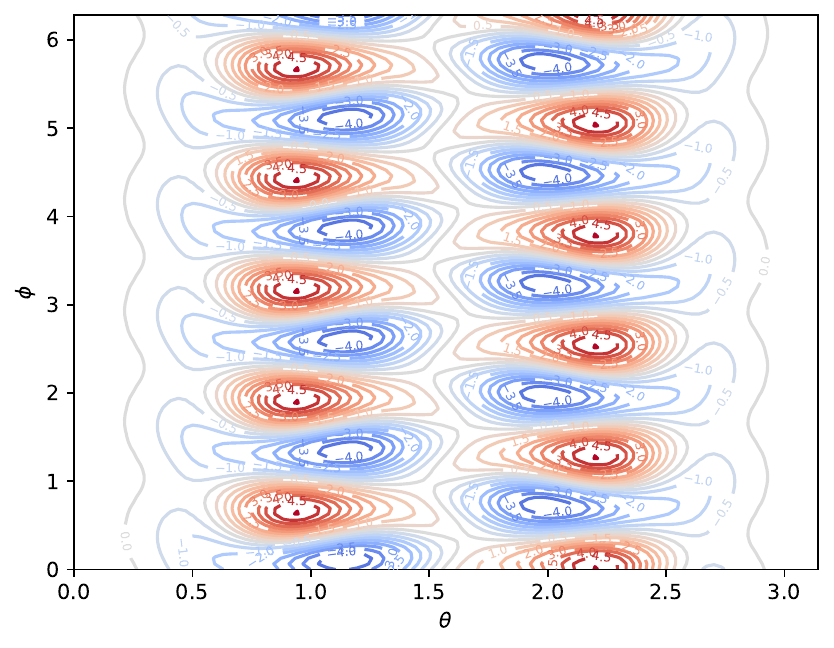}}\\
            \subfloat[RPA@PBE Radial]{\includegraphics[scale=0.6]{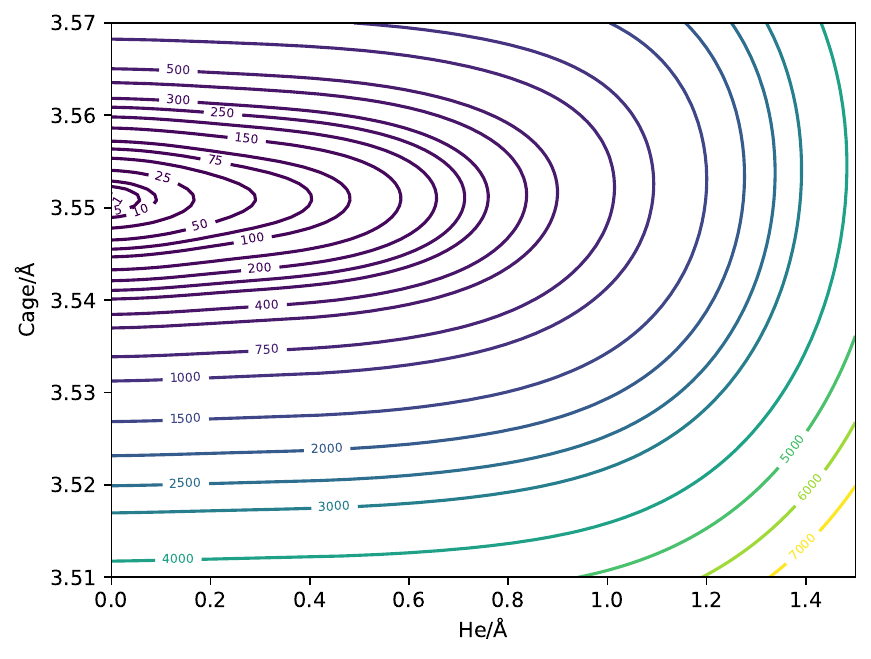}}\,
            \subfloat[RPA@PBE Angular]{\includegraphics[scale=0.6]{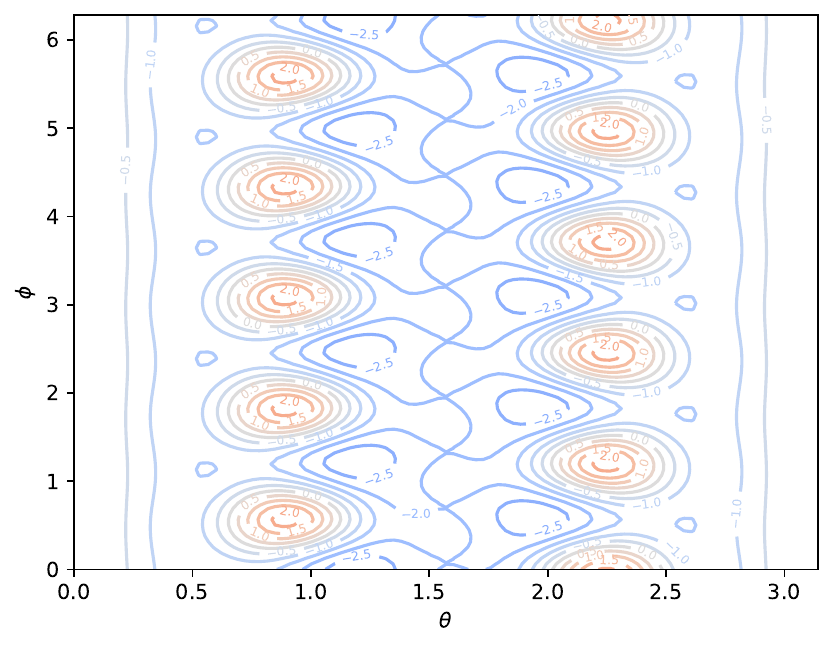}}
            \caption{Two dimensional (a) MP2 radial, (b) MP2 angular, (c) RPA@PBE radial, and (d) RPA@PBE angular slices of the PES. Radial slices are taken at $(\theta, \phi)=(0,0)$, the energy zero set to the minimum of the slice and contours at [1, 5, 10, 25, 50, 75, 100, 150, 200, 250, 300, 400, 500, 750, 1000, 1500, 2000, 2500, 3000, 4000, 5000, 6000, 7000]\wavenumber. Angular slices are taken at $(r_{\ce{He}}, R_{\ce{C60}})=(0.3066, 3.54181)$, the energy zero at the origin and contours are every 0.5\wavenumber in the interval [-5, 5]\wavenumber.}
            \label{fig: PES}
        \end{figure*}

        For the two-dimensional radial kernel, $k$ defined in Eq \eqref{eq: 2D kernel}, was chosen to take the form of an anisotropic Mat\'{e}rn covariance function with $\nu = 2.5$. For the four-dimensional corrections with kernel of the form given in Eq \eqref{eq: 4D kernel}, $k_r$ took the same form as $k$, whereas for $k_a$ we choose to use the Wendland-C4 covariance function shown in Table \ref{table: Spherical kernels}, with $\tau = 6$ and the great-circle metric defined in Eq \eqref{eq: Great Circle Distance} enforcing $s = \pi$. Both kernels required the optimisation of four hyperparameters namely: amplitude, $r_{\ce{He}}$ and $R_{\ce{C60}}$ length-scales and the noise. Further details for training the GPs to generate the PESs are outlined within the supplementary information, Section SI 2. 

        Adding the predictions of these kernels together gives rise to the two-dimensional radial and angular slices of the four-dimensional PES seen in Fig \ref{fig: PES}. As the PES is defined up to an arbitrary energy shift, we choose the energy zero to be the energy minimum in the radial slice which is taken at $(\theta, \phi) = (0, 0)$ and in the angular slice to be the value at the origin, with the radii frozen at $(r_{\ce{He}},\, R_{\ce{C60}})$ = (0.30066\AA, 3.54181\AA). 

        The shape of the radial slices is remarkably smooth along both radial coordinates. This indicates the radial surface can be well described by a polynomial oscillator. The $r_{\ce{He}}$ coordinate shows a considerable amount of anharmonicity as previously observed.\cite{bacanuExperimentalDeterminationInteraction2021c, jafariTerahertzSpectroscopyHelium2022} The $R_{\ce{C60}}$ coordinate on the other hand shows very little anharmonicity. As the elliptical contours are almost perfectly aligned along these axes, this suggests there is unlikely to be any observable coupling between the \ce{He} translational and \ce{C60} cage breathing modes.

        These features are shared in both the MP2 and RPA@PBE radial surfaces, alongside the equilibrium \ce{He} position being at the origin. This is expected given the $I_h$ symmetry of the system. However there is one stark difference being the equilibrium cage radius. MP2 predicts this to be at $R_{\ce{C60}}$ = 3.536\AA\, whereas the RPA@PBE predicts it to be at $R_{\ce{C60}}$ = 3.551\AA. This difference of approximately 0.04\% signifies the relevance of this coordinate in the surface which was previously neglected.\cite{bacanuExperimentalDeterminationInteraction2021c, jafariTerahertzSpectroscopyHelium2022} Fig \ref{fig: PES} also highlights the importance of the four-dimensional nature of the surface. Picking any particular one-dimensional slice through an \textit{ab initio} PES with fixed $(\theta, \phi, R_{\ce{C60}})$ will lead to different translational energies.

        Turning to the angular slices the contours are taken every 0.5\wavenumber in the interval [$-5$\wavenumber, $5$\wavenumber]. There is a great similarity between the positions of the positive four-dimensional corrections given in red, and the negative shown in blue. Both the MP2 and RPA@PBE surfaces have a periodic $\frac{2\pi}{5}$ symmetry along the $\phi$ direction as necessitated by the symmetrisation of the data with the $C_5$ rotation axis. They also exhibit the symmetry of mirroring along $\theta = \frac{\pi}{2}$ and swapping the $\phi = (0, \pi)$ and $\phi = (\pi, 2\pi)$ regions due to the inversion symmetry introduced into the training data.

         However, the numerical values of the corrections are subtly different between the MP2 and RPA@PBE. The MP2 corrections span the full corrections range, whereas the RPA@PBE, with its much lighter contours only spans the range [$-2.5$\wavenumber, $2$\wavenumber]. The sensitivity of these four-dimensional corrections is further illustrated in the SCS-MP2 and SOS-MP2 PES slices, which are given in the supplementary information Section SI 2. While these corrections are on the order of single digit wavenumbers, this weak angular dependence not being completely flat will lift the $(2l+1)$-fold degeneracy of the states in the traditional 3D HO. This is alongside the breaking of principal quantum number $n = 2k+l$ degeneracy due to the radial anharmonicity.

        \begin{figure*}
            \centering
            \setcounter{subfigure}{0}
            \subfloat[Radial lower bounds]{\includegraphics[scale=0.6]{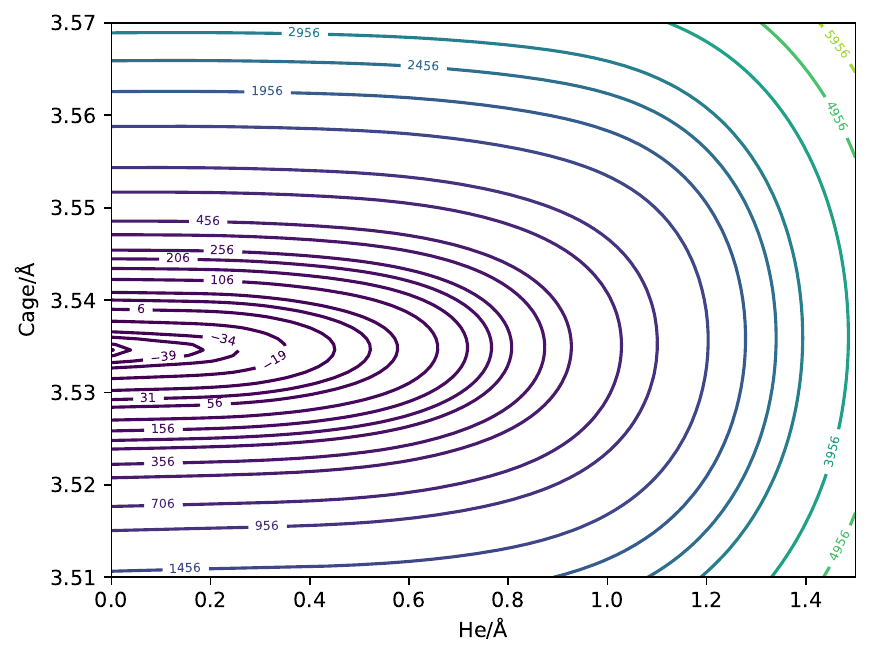}}
            \subfloat[Angular lower bounds]{\includegraphics[scale=0.6]{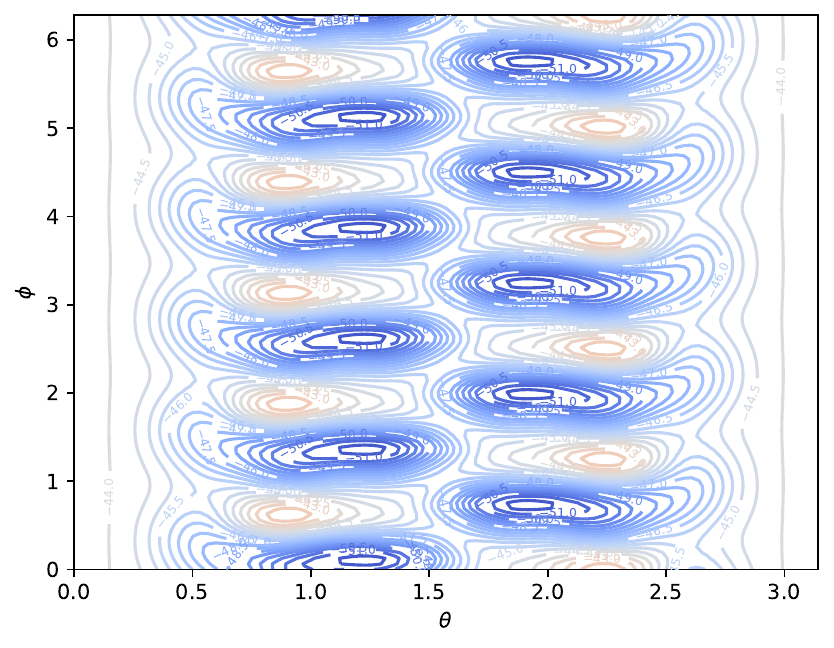}}\\
            \subfloat[Radial upper bounds]{\includegraphics[scale=0.6]{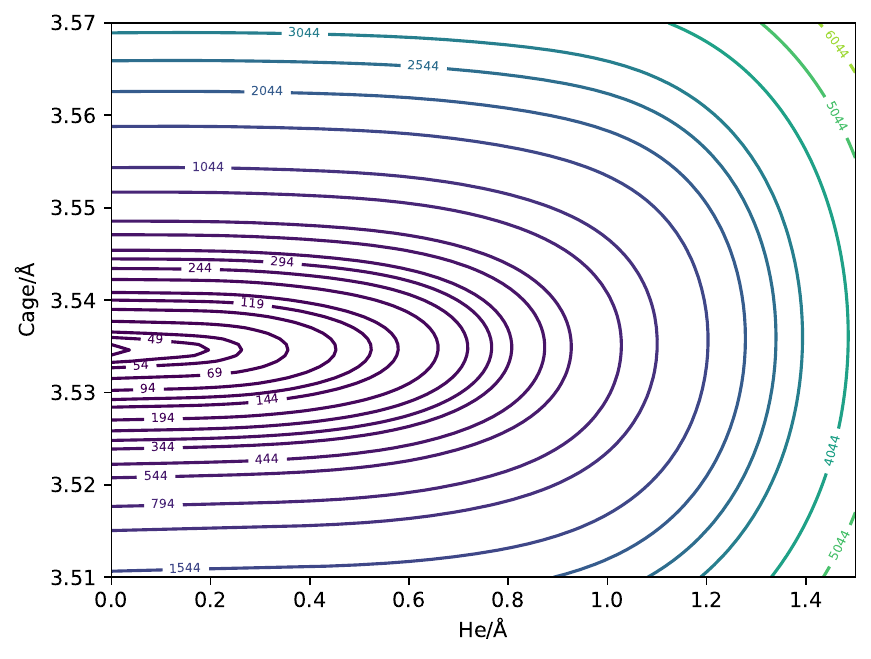}}
            \subfloat[Angular upper bounds]{\includegraphics[scale=0.6]{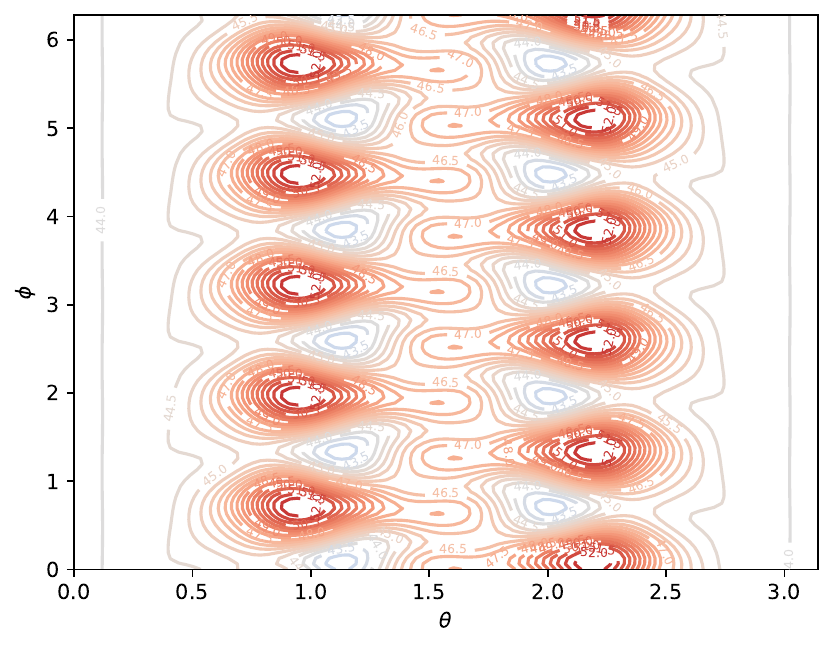}}
            \caption{95\% confidence interval on the value of the four-dimensional corrections to a fixed two-dimensional radial MP2 PES: (a) radial lower bound, (b) angular lower bound, (c) radial upper bound, and (d) angular upper bound. The radial and angular slices are at the same fixed coordinates as in Fig \ref{fig: PES}. Radial contours have their values shifted by -44\wavenumber in the lower bound, and +44\wavenumber in the upper bound. Angular contours are at every 0.5\wavenumber relative to the value at the origin being -43.5\wavenumber in the lower bound and +43.5\wavenumber in the upper bound.}
            \label{fig: PES confidence intervals}
        \end{figure*}

        The choice of angular kernel and metric was made by comparing the 95\% confidence interval on the surface prediction, whose lower and upper bound two-dimensional slices are shown in Fig \ref{fig: PES confidence intervals}. The radial slices are taken at the same fixed angular coordinates and vice versa as in Fig \ref{fig: PES}. These confidence intervals are calculated only from the error in the prediction of the four-dimensional corrections. This is equivalent to ``pre-selecting'' a two-dimensional radial surface and asking what is the possible range of values of the four-dimensional corrections. If the GPs are independent, the total error in each prediction would be given by $\sqrt{\Sigma_{2D}^2+\Sigma_{4D}^2}$. However, in this particular form of PES, this is likely to be an upper bound to the true confidence interval, as the two-dimensional GP and four-dimensional GP would have a negative cross-covariance. This is because as the two-dimensional surface becomes more accurate, the four-dimensional surface becomes more flat. 

        The radial lower and upper bound surfaces for MP2 look almost identical to the mean surface prediction. The contours are centred at the minimum of the surface, but taken at 44\wavenumber lower and higher than the values in Fig \ref{fig: PES}. While this confidence interval of 88\wavenumber may seem quite large, as the PES is only defined up to an arbitrary energy zero, it can be linearly shifted without affecting the subsequent calculations. This is an indication of the kernel constraining the behaviour of the GP, strongly enforcing the overall shape of the PES. As further calculations using the PES are agnostic of the energy zero, the mapping $E=E_{2D}+E_{4D}\rightarrow (E_{2D}+\delta E_{2D})+(E_{4D}+\delta E_{4D})$ will maintain the subsequent properties derived from the PES. Therefore this wide confidence interval is not diagnostic of a poor PES.

        The angular lower and upper bounds however, show a more involved behaviour. With the centre of the contours at the origin, at -43.5\wavenumber and +43.5\wavenumber respectively, the confidence interval still maintains the same $\frac{2\pi}{5}$ periodic and inversion symmetry. While the range of the upper and lower bound surfaces are still approximately 10\wavenumber, analogous to the mean prediction, there is a dominance towards the extremes of each bound. That is to say, the lower bound surface shows tendency to skew towards a more negative correction and vice-versa. This implies the confidence interval is somewhat spiky, with a width in places of over 100\wavenumber. Once again, as the PES is only defined up to an arbitrary energy zero, the shape of these errors is of more importance. The consequence of the exact shape and size of the PES and confidence interval will be observed in the degree of splitting of the $(2l+1)$-fold degenerate energy levels and the size of error bars on the translational eigenstates seen in Fig \ref{fig: Translational Energies}.

        This choice of kernel and metric had the tightest confidence interval of all possible options indicated in Table \ref{table: Spherical kernels}. The ranges of the other options of kernel and metric for MP2 are given in the supplementary information Section SI 2. While the shape of these confidence intervals and PES is more important than its width, it is still prudent to try to tighten them. The errors could arise due to a combination of factors including the choice of metric which dictates the value of the support parameter $s$ defined in Table \ref{table: Spherical kernels}, or even because $\tau$ is fixed. They may also be a consequence of the tight hyperparameter bounds on the four-dimensional kernel imposed by chemical intuition. However, widening them could lead to a very flat surface prediction by the GP. On the other hand, introducing more training data could lead to overfitting of the surface. Noticing the Wendland-C4 being an improvement on the Wendland-C2 opens a possible systematic route to better choices for $k_a$. This could be achieved by generating higher degree Wendland functions\cite{wendlandPiecewisePolynomialPositive1995} and using them with the great-circle metric as the angular kernel.

    \subsection{Translational Eigenstates \label{sec: Results Energies}}

        \begin{figure*}
            \centering
            \includegraphics[scale=1.2]{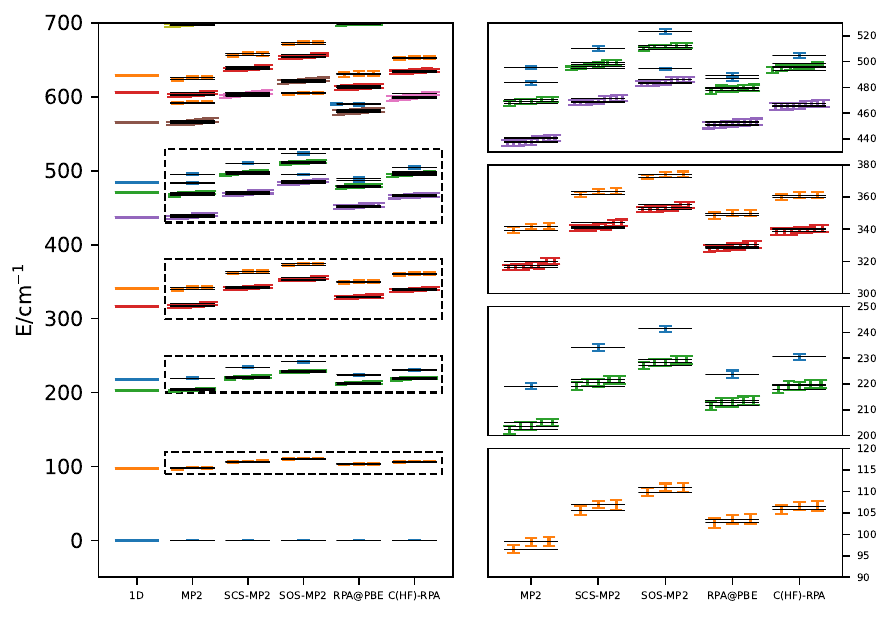}
            \caption{Four-dimensional translational energies of \Endo{^3He}{C60} for MP2, SCS-MP2, SOS-MP2, RPA@PBE, and C(HF)-RPA PESs alongside the one-dimensional experimental derived counterparts. One-dimensional energies are coloured by angular momentum quantum number $l$ in increasing order: blue, orange, green, red, purple, brown, pink and grey. Error bars on the electronic structure methods are also coloured according to the near degeneracy of states.}
            \label{fig: Translational Energies}
        \end{figure*}

        The nuclear Hamiltonian in Eq \eqref{eq: 4D Hamiltonian} was diagonalised using basis functions of the form defined in Eq \eqref{eq: 4D basis set}, with 10 translational $\ket{kl}$ functions for each $l$, spherical harmonics $\ket{lm}$ with $l\leq7$, and 4 cage breathing $\ket{b}$ functions. This gave an overall basis set size of 2560 functions. The 100 lowest lying eigenstates of \Endo{^3He}{C60} were converged to within 0.5\wavenumber, with the majority of states converging to within 0.02\wavenumber.

        These eigenstates are plotted, with the ground state energy set to be the energy zero, in Fig \ref{fig: Translational Energies}, alongside energies from the spherically symmetric one-dimensional potential. As these were generated from a potential fitted to the spectroscopic data, they are equivalent for comparison. These one-dimensional energies are colour coded by their angular momentum quantum number, $l$. Due to the spherical symmetry, these states are $(2l+1)$-fold degenerate, but the principal quantum number of the 3D HO $n = 2k+l$ which usually gives $\lceil\frac{n}{2}\rceil$-fold degeneracy (ignoring the $l$ degeneracy effect), is lifted. This is due to the anharmonic nature of the potential along $r_{\ce{He}}$, which is also seen in Fig \ref{fig: PES}.

        This anharmonicity is recovered in all five electronic structure methods: MP2, SCS-MP2, SOS-MP2 RPA@PBE, and C(HF)-RPA. Looking at the energies of these translational eigenstates, it is noticeable that they are consistently substantially overestimated by the two semi-empirical methods SCS-MP2 and SOS-MP2. The fundamental transition differs by approximately 10\wavenumber\, to the spectroscopic data, with higher energy states having sequentially worse agreement. The two \textit{ab initio} methods on the other hand perform considerably better, with MP2 predicting a fundamental frequency of $96.58\pm0.99$\wavenumber which has very good agreement to the spectroscopic value of 96.7\wavenumber.\cite{bacanuExperimentalDeterminationInteraction2021c, jafariTerahertzSpectroscopyHelium2022} The RPA@PBE consistently overestimates these energies by approximately 6\wavenumber. While its corrected variant, C(HF)-RPA, yields an equilibrium cage radius significantly closer to that obtained with the MP2 method (see Section SI 2), the accuracy in terms of the energies of the translational eigenstates is deteriorated. C(HF)-RPA performs very similar to SCS-MP2 at low energy states, but as this energy increases it starts to outperform the semi-empirical methods. The possible causes of the discrepancies of MP2 and RPA to the one-dimensional data are described in Section \ref{sec: Theory ES}: the MP2 lacking screening effects, and the RPA@PBE not including exchange effects. While \textit{a priori} it is not obvious which is the more significant effect, the comparison to the spectroscopic data indicates that the MP2 outperforms the RPA@PBE. However, it should be noted that, due to the neglect of exchange effects between particle-hole pairs, RPA@PBE exhibits significantly higher efficiency compared to the MP2 method. To illustrate, a single MP2 calculation using a quadruple-zeta basis set takes approximately 325\,min, whereas the corresponding RPA@PBE calculation requires about 34\,min. This makes the MP2 method roughly ten times as computationally expensive as the RPA calculation. Therefore, although the energies predicted by RPA@PBE do not converge to spectroscopic accuracy, the significantly higher computational efficiency of this electronic structure method suggests that it should not be discarded as a viable choice for these systems.
        
        We also observe a lifting of the of the $(2l+1)$-fold degeneracy in all four electronic structure methods, but to varying degrees, indicating different levels of importance on angular dependence within the PES. All three MP2 methods predict a $p$-state splitting of over 1\wavenumber and a $d$-state splitting ranging to over 3\wavenumber whereas the RPA@PBE and C(HF)-RPA predict these splittings to be less than 1\wavenumber and 2\wavenumber respectively. However, considering the $I_h$ character table, all three $p$ states and five $d$ states are expected to be degenerate. This fictitious splitting arises due to the symmetrisation of the data using only a single $C_5$ rotation axis and the centre of inversion $i$. This only generates a subgroup symmetry of $I_h$, namely $D_{5d}$, where the $p$ and $d$ state splitting is expected. Fully symmetrising the training data is a non-trivial task due to the orientation of the \ce{C60} cage shown in Fig \ref{fig: C60} and position of the remaining symmetry elements. This would also be counter-productive to the problem as it would lead to strong overfitting in the PES, making it very spiky at the location of the training data.
        
        For states with $l\geq 3$, namely $f$ states and higher, there is a real splitting of these states as expected under $I_h$ symmetry, of 4\wavenumber in the MP2 type and 2.5\wavenumber in the RPA type methods. The smaller splitting is expected in the RPA, as the range of the angular PES slice in Fig \ref{fig: PES} is smaller than the equivalent MP2 range. This real splitting indicates the importance of catering for the angular dependence in the PES, as this allows for states with differing $l$ to couple and mix, which is not possible with a purely symmetric potential. The lifting of this degeneracy may be masked within an experimental spectrum into the linewidth of the corresponding transition.

        For the states with energies under 500\wavenumber, there is a one-to-one mapping of the one-dimensional energy levels, to the ones of each electronic structure method. However, around 500\wavenumber, an extra state appears in the electronic structure diagonalisation. This is due to the excitation of a quantum in the cage breathing mode but remaining in the ground translational state. In the three MP2 methods, the fundamental cage breathing frequency is 495\wavenumber, but 490\wavenumber in the RPA@PBE. Looking at the top right insert in Fig \ref{fig: Translational Energies}, this extra state is obvious in the MP2, SOS-MP2 and RPA@PBE, but seems to be missing in the SCS-MP2 and C(HF)-RPA. It is necessarily still present, but has been mixed into the cluster of the first set of excited $d$ states. 

        Going upwards in energy, all the states with energy $<500$\wavenumber will be repeated again, with an excitation in the breathing mode. These will be combined with the higher pure translational excitations which leads to a cluttered energy level diagram and increases the difficulty in matching the one-dimensional energies to the electronic structure analogues. There seems to be no discernible coupling of the breathing mode to the translational mode, as the combination modes occur at the sum of energies of exciting each mode separately. This is expected given the form of radial slices in the PES in Fig \ref{fig: PES}, with the contours aligning almost perfectly along the radial axes.

        An advantage of using GPR for PES interpolation, is the ability to generate the covariance matrix as defined in Eq \eqref{eq: GP covariance}, alongside its mean prediction given by Eq \eqref{eq: GP mean}. We construct the overall covariance matrix as $\boldsymbol{\Sigma} = \boldsymbol{\Sigma}_{2D,2D}+\boldsymbol{\Sigma}_{4D,4D}$, which is exact assuming independence between each GP. However, as mentioned in Section \ref{sec: Results PES}, there is likely to be a negative cross covariance between these processes denoted by $\boldsymbol{\Sigma}_{2D,4D}$, implying the error bars shown in Fig \ref{fig: Translational Energies} are upper bounds to the ``true'' errors. These error bars were calculated by generating 100 samples of the PES from the appropriate multivariate Gaussian distribution which in turn generates 100 different nuclear Hamiltonian matrices. The eigenvalues of these matrices were averaged, generating its mean and standard deviation. These error bars are also colour coded by near degeneracy of the states. For states with energies below 500\wavenumber, this corresponds identically to the angular momentum quantum number, $l$, with the spectroscopic reference. Above this, however, this may not be the case due to the muddling of states involving an excitation in the breathing mode. 

        Looking at the four zoomed-in inserts in Fig \ref{fig: Translational Energies}, it is noticeable how small these error bars are, usually less than $\pm2$\wavenumber. Despite the wide looking 95\% confidence interval seen in Fig \ref{fig: PES confidence intervals}, this ends up being a relatively small error bar in the translational energy, which may be interpreted in relation to a linewidth in an experimental spectrum. The minute size of these error bars once again demonstrates the dominance of the two-dimensional radial PES, compared to the four-dimensional corrections. The error bars on the MP2 translational energies are smaller than those of the RPA@PBE, despite the angular corrections being larger in magnitude. This implies the confidence interval on the MP2 surface is tighter than the respective one for RPA@PBE.

        \begin{figure*}
            \centering
            \setcounter{subfigure}{0}
            \subfloat[$s$]{\includegraphics[scale=0.23]{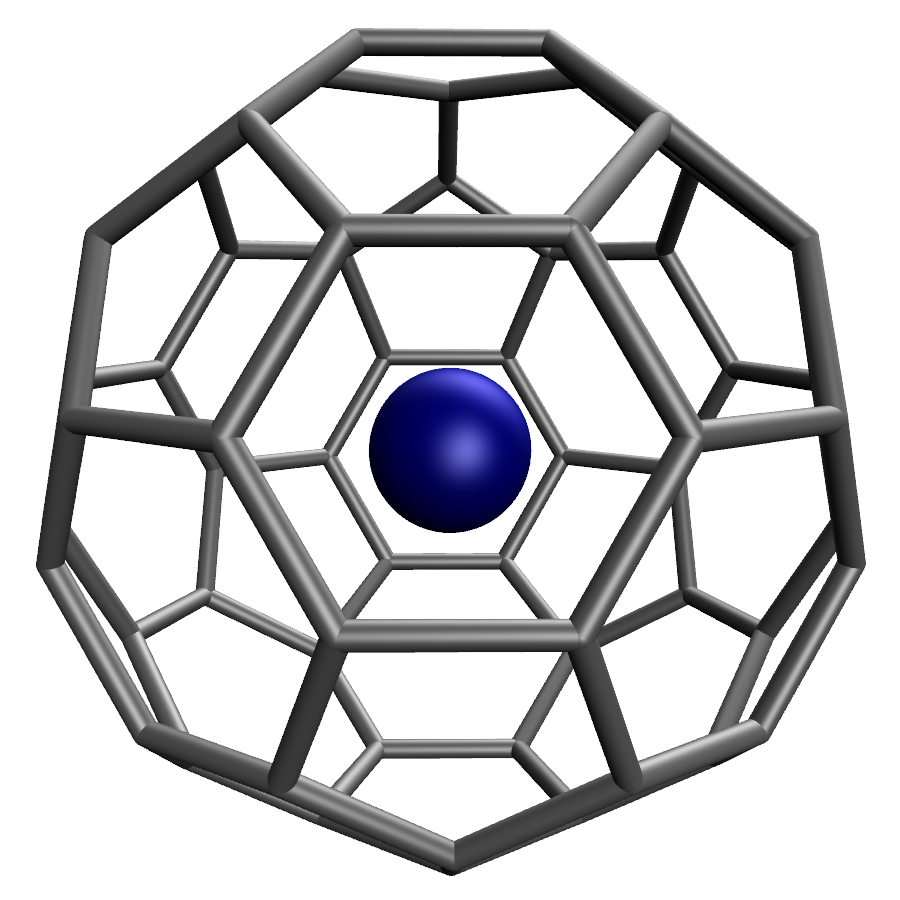}}\,
            \subfloat[$p$]{\includegraphics[scale=0.23]{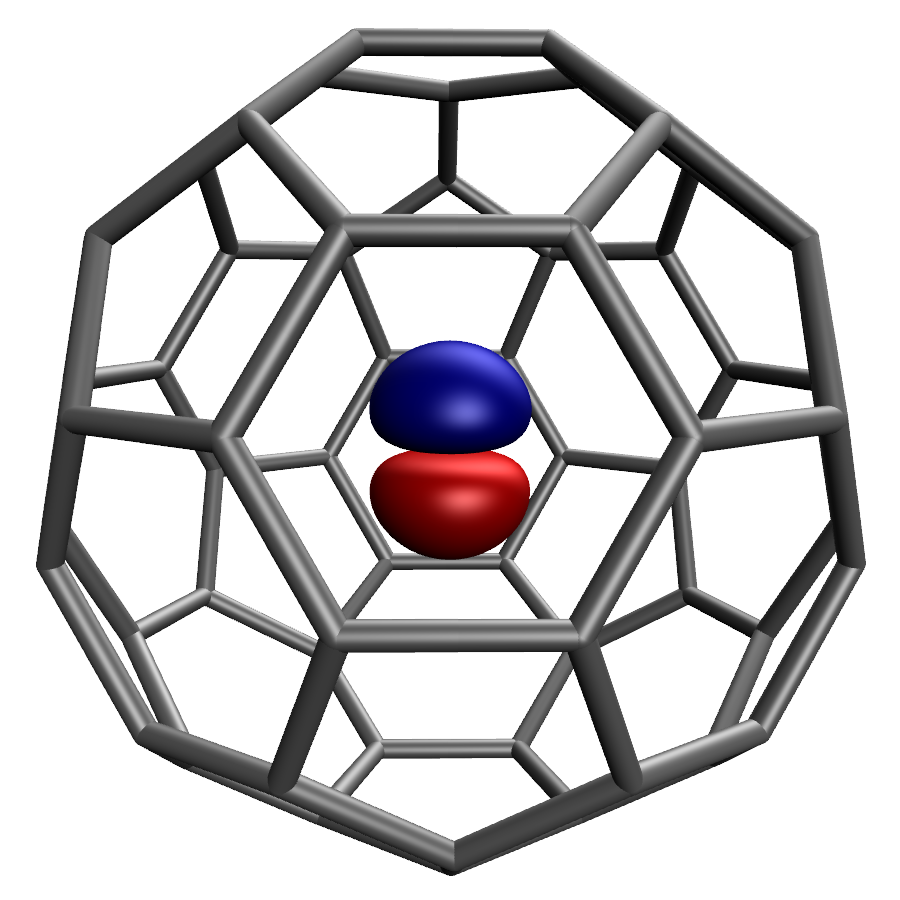}}\,
            \subfloat[$d$]{\includegraphics[scale=0.23]{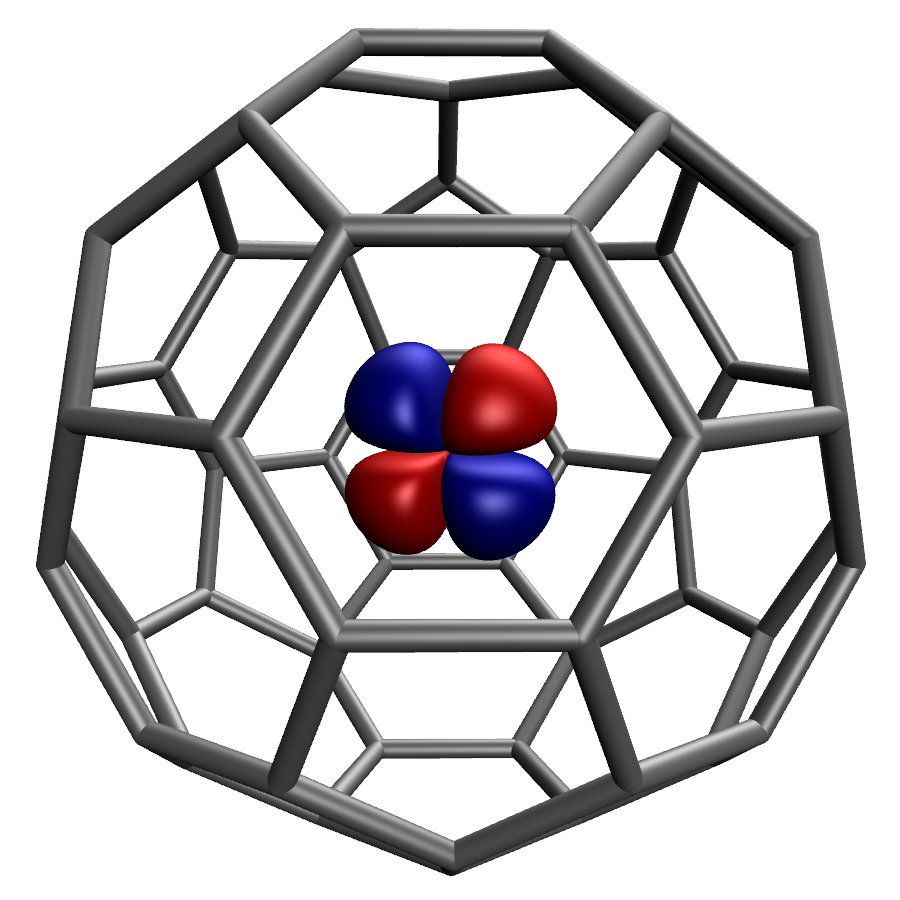}}\,
            \subfloat[$f$]{\includegraphics[scale=0.23]{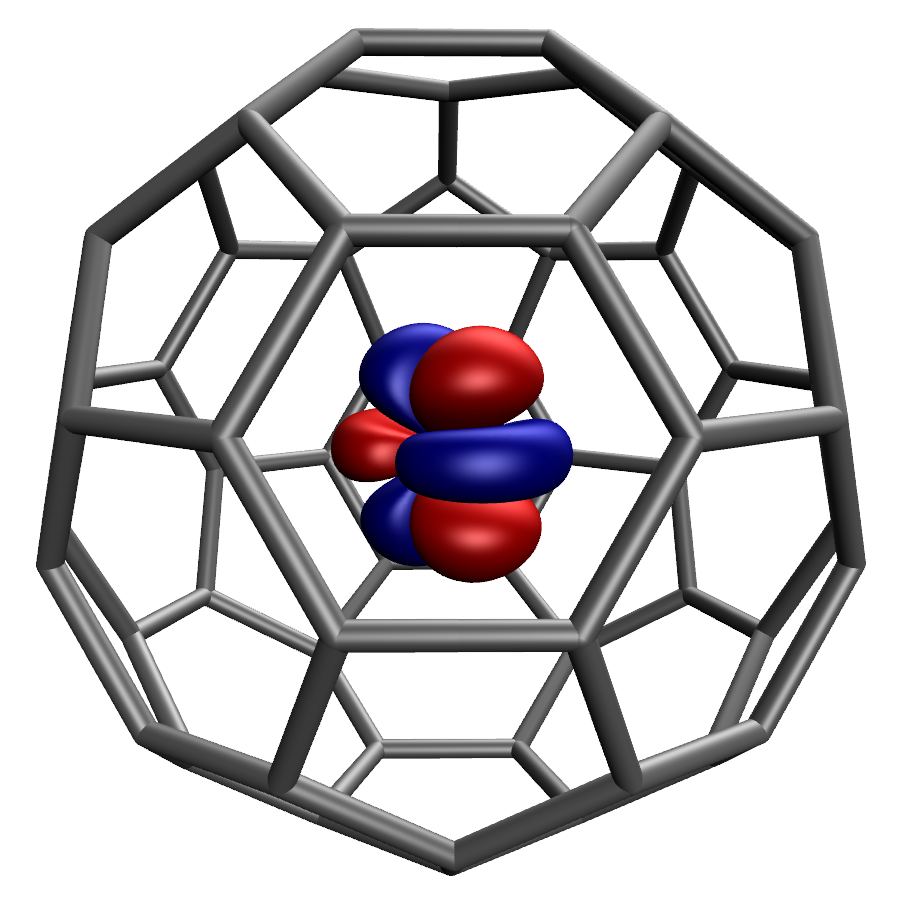}}
            \caption{MP2 isosurfaces of (one of) the lowest lying set of (a) $s$, (b) $p$, (c) $d$, and (d) $f$ nuclear orbitals. The \ce{C60} cage has been rotated relative to Fig \ref{fig: C60}, with the $z$ direction pointing directly upwards, and $xy$ plane rotated to centre down a $C_3$ axis. The isosurfaces are taken at 0.5\% of the maximum amplitude of each wavefunction.}
            \label{fig: Wavefunctions}
        \end{figure*}

        \begin{table}
            \centering
            \begin{tabular}{ccc}
                 \hline Electronic Structure Method & Overlap & Hellinger Distance  \\
                 \hline MP2&0.99290&0.08424\\
                 SCS-MP2&0.98675&0.11508\\
                 SOS-MP2&0.98288&0.13081\\
                 RPA@PBE&0.98885&0.10560\\
                 C(HF)-RPA&0.98675&0.11511\\
                 \hline
            \end{tabular}
            \caption{Overlap and Hellinger distance of the ground state wavefunction for MP2, SCS-MP2, SOS-MP2, RPA@PBE, and C(HF)-RPA with the experimentally derived one-dimensional ground state.}
            \label{table: Hellinger ground state}
        \end{table}

        \begin{table}
            \centering
            \begin{tabular}{c|ccccc}
                 $\bra{\downarrow}$, $\ket{\rightarrow}$& MP2 & SCS-MP2 & SOS-MP2 & RPA@PBE&C(HF)-RPA  \\
                 \hline MP2&-&0.04055&0.05880&0.02617&0.03856\\
                 SCS-MP2&0.04055&-&0.01831&0.02357&0.00901\\
                 SOS-MP2&0.05880&0.01831&-&0.03987&0.02311\\
                 RPA@PBE&0.02617&0.02357&0.03987&-&0.01782\\
                  C(HF)-RPA&0.03856&0.00901&0.02311&0.01782&-
            \end{tabular}
            \caption{Hellinger distances of the ground state between all five electronic structure methods. Diagonal elements, by definition, are necessarily zero.}
            \label{table: Hellinger ES}
        \end{table}

        The energies are not the only information obtained from the diagonalisation procedure; the eigenvectors corresponding to the wavefunctions are also obtained. These nuclear wavefunctions (orbitals) have strikingly regular patterns as seen in Fig \ref{fig: Wavefunctions}, which allows for simple assignment of both translational and total angular momentum quantum numbers $k$ and $l$ respectively. These eigenfunctions are usually dominated by a single 3D HO wavefunction. With the reorientation of the cage to have the $z$ direction pointing upwards, it is interesting to see the rotation along this axis of the $f$ orbital in relation to the $d$ orbital. It turns out neither are perfectly aligned along either the $x$ or $y$ directions, suggesting that this particular set of $x$ and $y$ axes is not best suited to the Cartesian representation of these orbitals. The isosurfaces, taken at 0.5\% of the maximum amplitude of each wavefunction are constrained within a sphere of radius 1.5\AA\, which is less than 50\% the value of the \ce{C60} cage radius. This implies that while the \ce{He} is encapsulated within the \ce{C60} cage, it does not explore any region close to the \ce{C} atoms.

        Instead of using the energies as the discriminator for the validity of electronic structure method, the wavefunction can be used instead. This is achieved by considering the overlap, or by calculating the Hellinger distance, as defined in Eqns \eqref{eq: 4D overlap} and \eqref{eq: Hellinger distance} respectively between the one-dimensional reference state and four-dimensional eigenfunction. Considering just the ground state, we take the reference $\bra{LM}\bra{B}$ states to be the pure basis state $\bra{00}\bra{0}$, reducing the summation over $lmb$ quantum numbers to a single term. Due to the different equilibrium cage radii predicted by each electronic structure method, the centre of expansion of these breathing 1D HO functions is different. For each electronic structure method, we enforce the reference 1D breathing state to be centred in the same position. This has the intended effect of the Hellinger distance (and overlap) to be purely due to the difference in the translational portion of the eigenfunction. These overlaps and distances are given in Table \ref{table: Hellinger ground state}. While all these methods have over 98\% overlap with the reference one-dimensional ground state, the distinction in their quality is more apparent in their Hellinger distances. In line with the energy matches, the MP2 and RPA@PBE perform best, with the C(HF)-RPA and semi-empirical SCS-MP2 and SOS-MP2 lagging behind. This seems to indicate that either the energy, or the Hellinger distance could be used interchangeably as the validator of the electronic structure method. 
        
        Despite the spectroscopic accuracy in the energy prediction of the MP2, the Hellinger distance still seems fairly large. This could be due to the extra angular dependence in the potential, allowing the mixing of states with differing angular momentum quantum number $l$ to couple, which is not possible in a purely spherically symmetric potential. This casts doubt on the correctness of the one-dimensional potential as while this fits the spectroscopic data, might not correspond to any particular one-dimensional slice of the four-dimensional PES.

        Not only can the wavefunctions calculated from the electronic structure methods be compared to just the one-dimensional reference, they can also be compared to each other. The Hellinger distances of the ground state between electronic structure methods is given in Table \ref{table: Hellinger ES}. Once again, for the $\braket{B|b}$ integral, despite the differing centres of expansion, we enforce this to be $\delta_{Bb}$ to isolate just the translational states. Due to the high frequency of the cage breathing mode, if these integrals were left in, the distances would all be over 0.70. Considering just the translational states, we find that all these methods are closer to each other than they are to the one-dimensional reference. This is possibly due to the angular dependence, with the sum over $lmb$ in Eq \eqref{eq: Hellinger distance} not collapsing as both wavefunctions are four-dimensional. The SCS-MP2 and SOS-MP2 are most similar, closely followed by the MP2 and RPA@PBE, which follows suit from the energy level diagram in Fig \ref{fig: Translational Energies}. As with the energies, the C(HF)-RPA sits in between these two pairs, but falling closer to the SCS-MP2.

    \section{Conclusion \label{sec:Conc}}
    	In this work, we have investigated how the angular dependence of the potential energy surface and the cage breathing mode impact the translational eigenstates of the \Endo{He}{C60} endofullerene. For this investigation, we calculated complete basis set extrapolated four-dimensional potential energy surfaces, incorporating three He translational degrees of freedom and the \ce{C60} cage radius, using various electronic structure methods: MP2, SOS-MP2, SCS-MP2, RPA@PBE, and C(HF)-RPA. The rationale for utilising this array of methods is twofold: by comparing their results, we have gained not only confidence in our findings but also valuable insights into the performance of these methods where comparisons with reference data were feasible.

    Due to the high computational cost of the electronic structure calculations, the full Potential Energy Surface (PES) required interpolation from this sparse training data which was performed using Gaussian Process Regression. The PES was partitioned into a two-dimensional radial surface, to which corrections were applied making it overall four-dimensional. This is a well motivated choice due to the lack of obvious angular splitting within the spectroscopic data. The expression of the PES using spherical polar coordinates motivated the form of kernels shown in Eqns \eqref{eq: 2D kernel} and \eqref{eq: 4D kernel}, which required a nuanced choice of the angular kernel $k_a$ and its metric. We chose to use the Wendland-C4, as shown in Table \ref{table: Spherical kernels} with the great-circle metric as defined in Eq \eqref{eq: Great Circle Distance} as this accurately reflected the symmetry within the endofullerene, and had the smallest 95\% confidence interval as seen in Figs \ref{fig: PES}. and \ref{fig: PES confidence intervals}. 

    We generated the four-dimensional translational eigenstates of \Endo{He}{C60} by diagonalising the nuclear Hamiltonian. By not enforcing spherical symmetry in the potential, and allowing states for different angular momenta to couple we find the $(2l+1)$-fold degeneracy of states to be lifted. The three MP2 type methods all predict very similar levels of splitting, whereas the RPA methods predict smaller splittings. While some of this is artificial due to not exploiting the full $I_h$ symmetry of the system, we do observe a true splitting of $f$ states and higher of approximately 4\wavenumber. This suggests while the angular dependence is weak, it can still be observable in the broadening of these peaks in the spectra. On the other hand, we find very little influence on the cage breathing mode on these translational states.
    
    Comparing the energies to the one-dimensional case, we find that the RPA@PBE and MP2 outperform the C(HF)-RPA and the semi-empirical methods SCS-MP2 and SOS-MP2. Comparing the energies as seen in Figure \ref{fig: Translational Energies}, MP2 has the best agreement with the one-dimensional case, achieving sub-wavenumber accuracy with the spectroscopic data.\cite{bacanuExperimentalDeterminationInteraction2021c, jafariTerahertzSpectroscopyHelium2022} On the other hand, RPA@PBE exhibits a discrepancy of approximately 6\wavenumber compared to the experimental results; however, it proves significantly more efficient than MP2, surpassing it by almost an order of magnitude.

    Furthermore, the wavefunctions of the ground state were compared to the one-dimensional case through their overlap and Hellinger distances in Tables \ref{table: Hellinger ground state} and \ref{table: Hellinger ES}. Once again we find that the MP2 and RPA@PBE eclipse the C(HF)-RPA and the semi-empirical SCS-MP2 and SOS-MP2 methods. Considering the overall efficacy of these methods, and the trade-off between accuracy and computational cost we can recommend both MP2 and RPA@PBE as viable electronic structure methods for these systems.

    Going forwards, a natural extension would be to apply these electronic structure calculations to larger endofullerenes, whether that be molecules within \ce{C60} such as \Endo{H2}{C60}\cite{xuH2HDD22008, xuCoupledTranslationrotationEigenstates2009, xuInelasticNeutronScattering2013, xuQuantumDynamicsCoupled2008} and \Endo{H2O}{C60}\cite{bacicCoupledTranslationRotation2018, rashedInteractionsWaterMolecule2019, carrillo-bohorquezEncapsulationWaterMolecule2021, xuFullerene60Inelastic2022} which are popular in the literature or by considering larger fullerene cages such as \Endo{Ne}{C70}.\cite{mandziukQuantumThreeDimensional1994} An extra challenge for the latter system is to accurately describe the double well that is opened up by elongating the cage along the unique axis and reducing the symmetry to $D_{5h}$.\cite{panchagnulaExploringParameterSpace2023b}

    While knowledge of the PES gave us access to the translational eigenstates, in order to accurately reproduce an experimental spectrum the knowledge of intensities of transitions is also required. This necessitates the calculation of a dipole moment surface (DMS), which could be generated and interpolated in an analogous way to the PES\cite{kaluginaPotentialEnergyDipole2017}. Knowledge of the DMS allows for the calculation of the transition dipole moment integrals which are fundamental in calculating the transition intensities.

    \section*{Supplementary Information}
    	See the supplementary information for more details on the electronic structure methods used, the training of the Gaussian Process and diagonalisation procedure. Extra results using different electronic structure methods, kernels and metrics are also provided.
    
    \begin{acknowledgements}
        D.~G.~acknowledges funding by the Deutsche Forschungsgemeinschaft (DFG, German Research Foundation) -- 498448112. D.~G.~thanks J.~Kussmann (LMU Munich) for providing a development version of the FermiONs++ software package.
    \end{acknowledgements}

    \section*{Data Availability Statement}
    The data that support the findings of this study are openly available in Apollo - University of Cambridge Repository at https://doi.org/10.17863/CAM.105420\cite{KripaPanchagnulaDataset2023}, reference number \citenum{KripaPanchagnulaDataset2023}.

    \bibliography{HeC604D}

\end{document}


\title[SI: \Endo{He}{C60} 4D]{Supplementary Information: Translational eigenstates of \Endo{He}{C60} from four-dimensional \textit{ab initio} Potential Energy Surfaces interpolated using Gaussian Process Regression}
    \author{K. Panchagnula*\orcidlink{0009-0004-3952-073X}}
    \email{ksp31@cam.ac.uk}
     \affiliation{Yusuf Hamied Department of Chemistry, University of Cambridge, Cambridge, United Kingdom}
    \author{D. Graf \orcidlink{0000-0002-7640-4162}}
    \author{F.E.A. Albertani}
    \author{A.J.W. Thom \orcidlink{0000-0002-2417-7869}}%
    
    \date{\today}

    \maketitle

    \section{Electronic Structure Calculations \label{secSI: ES}}
        All calculations were performed using a development version of the FermiONs++ program package, developed by the Ochsenfeld group.\cite{kussmannPreselectiveScreeningMatrix2013, kussmannPreselectiveScreeningLinearScaling2015, kussmannHybridCPUGPU2017} The calculations were executed on compute nodes equipped with 2 AMD EPYC 7302 CPUs, featuring in total 32 cores and 64 threads at 3.00 GHz. All reported runtimes are wall times, not CPU times.
        
        The evaluation of the exchange-correlation terms was carried out using the multi-grids defined in Ref.~\citenum{laquaImprovedMolecularPartitioning2018}, utilising a smaller grid during the SCF optimisation and a larger grid for the final energy evaluation. These grids were generated using the modified Becke weighting scheme.\cite{laquaImprovedMolecularPartitioning2018} The SCF convergence threshold was set to $10^{-7}$, both for the change in energy and for the commutator norm 
        $|| \mathbf{FPS} - \mathbf{SPF} ||$.
        
        We employ the integral-direct resolution-of-the-identity Coulomb (RI-J) method, as described by Kussmann et al.\cite{kussmannHighlyEfficientResolutionofIdentity2021}, for evaluating the Coulomb matrices. Additionally, for the evaluation of the exact exchange matrices, we utilise the linear-scaling semi-numerical exact exchange (sn-LinK) method developed by Laqua et al.\cite{laquaHighlyEfficientLinearScaling2020}
        
        All correlation methods employed in this study utilise the frozen-core approximation. For both the $\omega$-CDGD-RPA and the $\omega$-RI-CDD-MP2 methods, we set the attenuation parameter to $\omega = 0.1$. In the RPA calculations, the integration along the imaginary frequency axis was conducted using an optimised minimax grid,\cite{kaltakLowScalingAlgorithms2014, grafAccurateEfficientParallel2018} consisting of 15 quadrature points. For the MP2 calculations, we employed 7 Laplace points. Additionally, the following thresholds were selected for the MP2 calculations: 
        $\theta_{\text{NB}} = 10^{-8}$, $\theta^{iaP}_{\text{NB}} = 10^{-10}$,
        $\theta_{\text{3c}} = 10^{-8}$, $\theta_{\underline{ij}} = 10^{-9}$,
        $\theta_{\text{schwarz}} = 10^{-12}$. 
        For detailed descriptions of these thresholds and approximations, the reader is referred to the original publications (Ref.~\citenum{grafAccurateEfficientParallel2018} and Ref.~\citenum{glasbrennerEfficientReducedScalingSecondOrder2020}).
        
        All calculations were performed with Dunning's correlation-consistent polarised triple-$\zeta$ (cc-pVTZ), quadruple-$\zeta$ (cc-pVQZ), and quintuple-$\zeta$ (cc-pV5Z) valence basis sets\cite{balabanovSystematicallyConvergentBasis2005, dunningGaussianBasisSets1989, koputInitioPotentialEnergy2002, prascherGaussianBasisSets2011, wilsonGaussianBasisSets1999, woonGaussianBasisSets1993, woonGaussianBasisSets1994} along with their corresponding RI basis sets.\cite{weigendEfficientUseCorrelation2002, hattigOptimizationAuxiliaryBasis2005, brossExplicitlyCorrelatedComposite2013}
                    
    \section{Gaussian Process Potential Energy Surface \label{secSI: GP}}

        \begin{figure*}
            \centering
            \setcounter{subfigure}{0}
            \subfloat[SCS-MP2 Radial]{\includegraphics[scale=0.6]{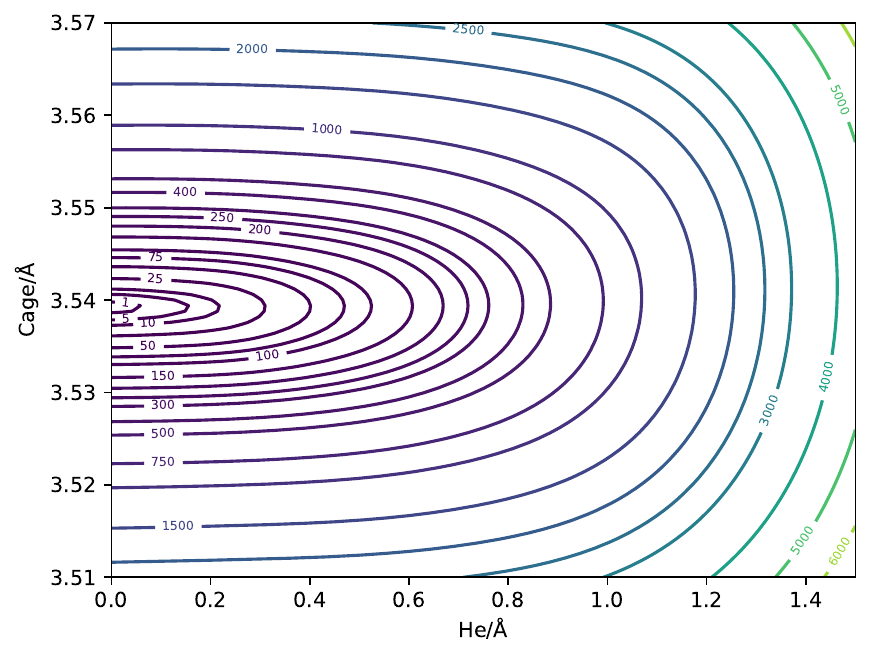}}\,
            \subfloat[SCS-MP2 Angular]{\includegraphics[scale=0.6]{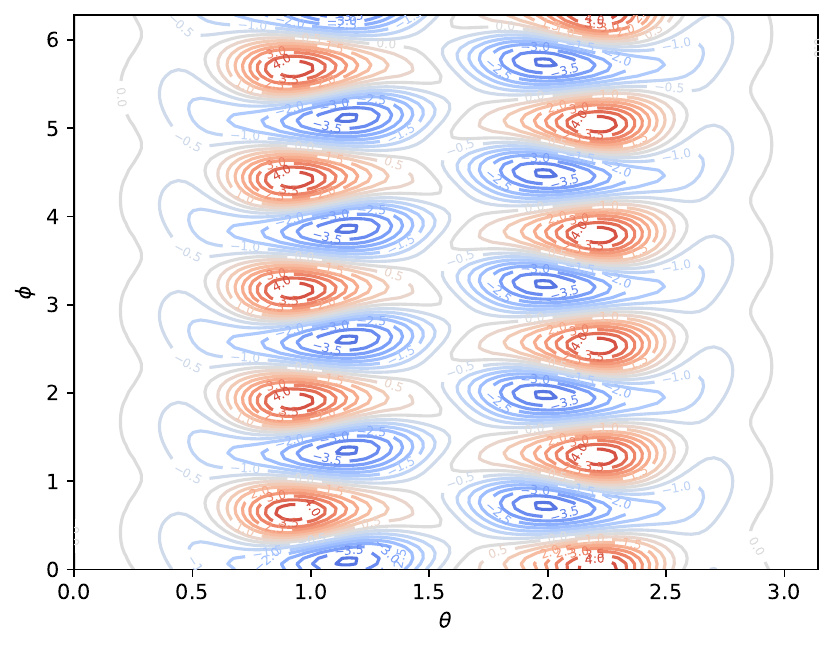}}\\
            \subfloat[SOS-MP2 Radial]{\includegraphics[scale=0.6]{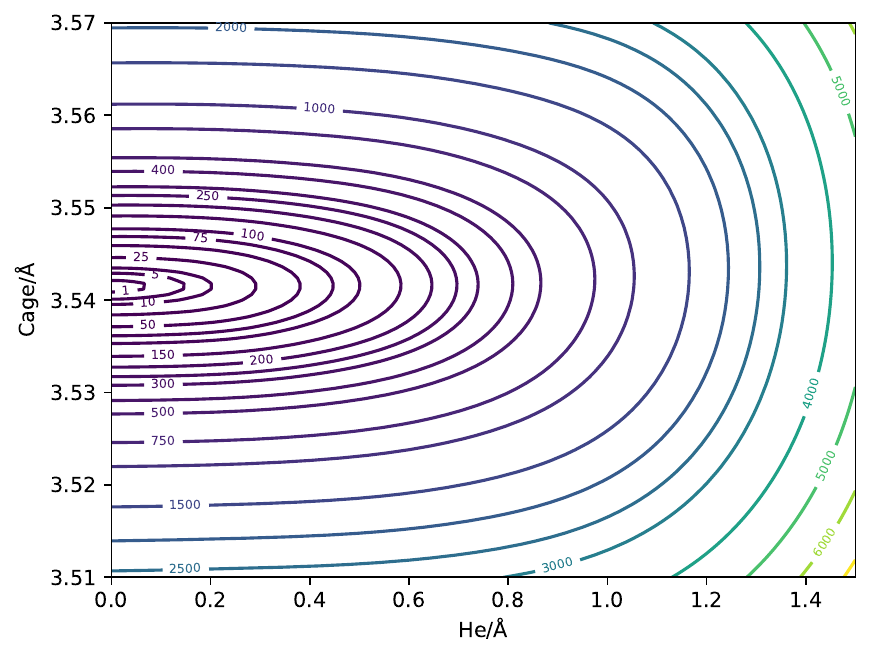}}\,
            \subfloat[SOS-MP2 Angular]{\includegraphics[scale=0.6]{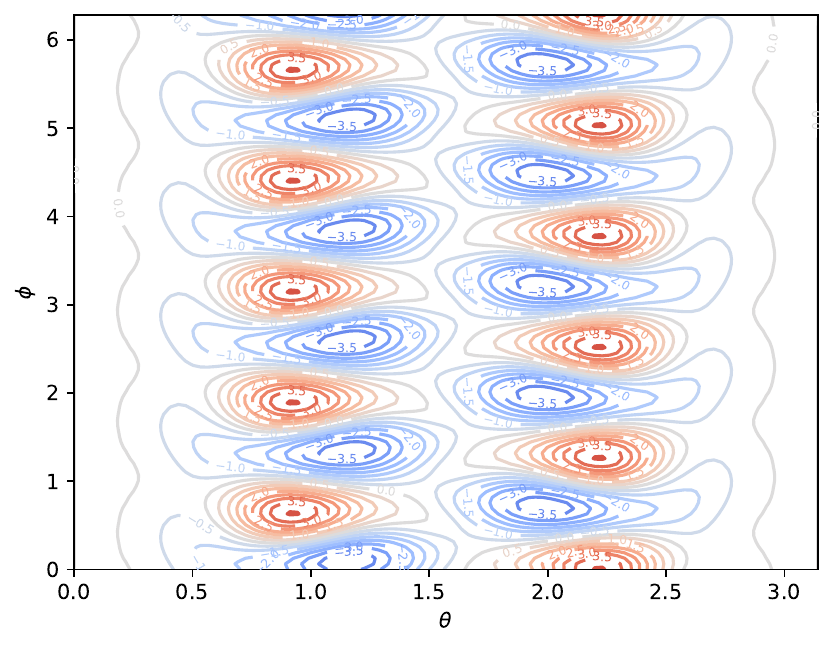}}\\
            \subfloat[C(HF)-RPA Radial]{\includegraphics[scale=0.6]{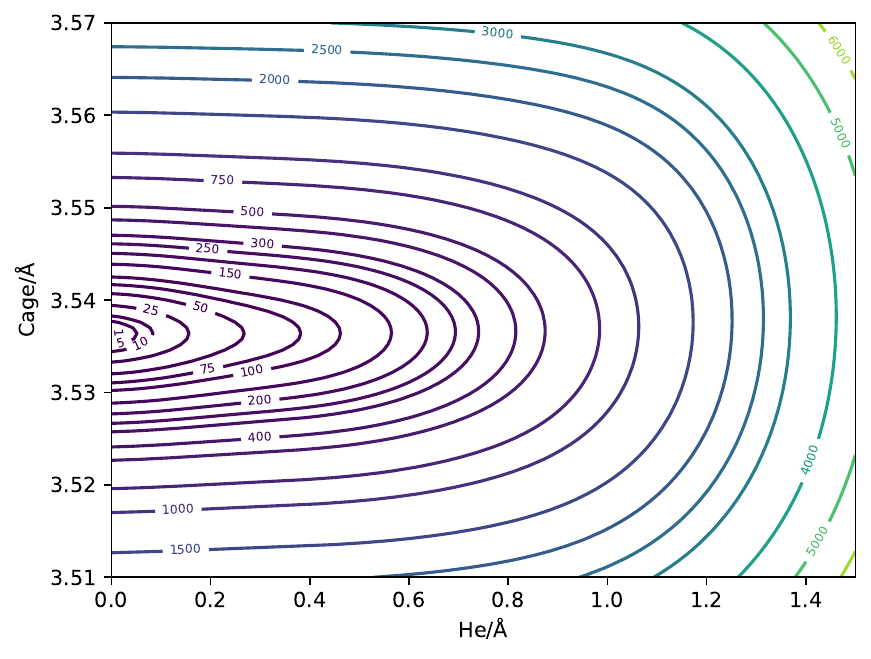}}\,
            \subfloat[C(HF)-RPA Angular]{\includegraphics[scale=0.6]{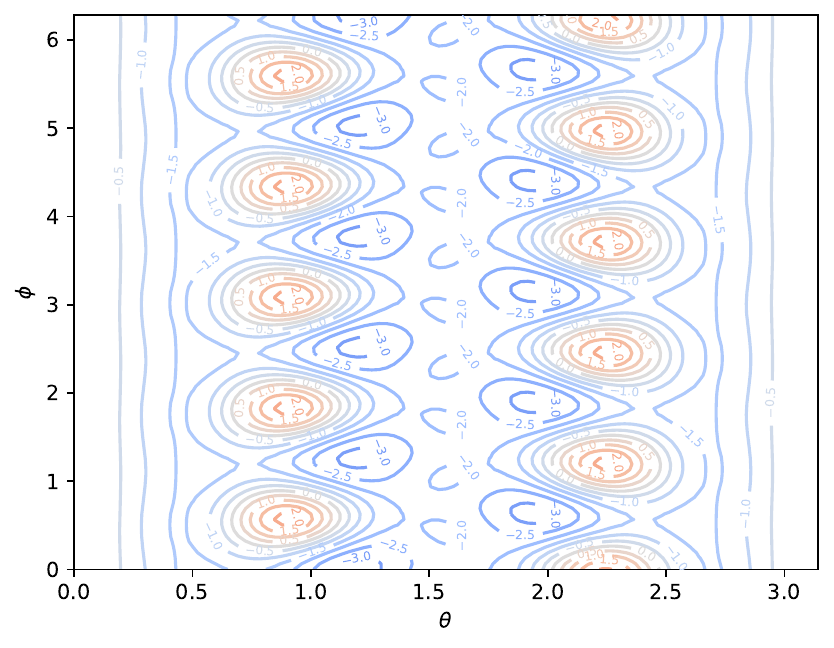}}
            \caption{Two dimensional (a) SCS-MP2 radial, (b) SCS-MP2 angular, (c) SOS-MP2 radial, (d) SOS-MP2, (e) C(HF)-RPA radial, and (f) C(HF)-RPA angular slices of the PES. Contour values, fixed angular coordinates for radial slices and fixed radial coordinates for angular slices are the same as Fig 2.
            }
            \label{fig: SCS SOS PES}
        \end{figure*}
        \begin{figure*}
            \centering
            \setcounter{subfigure}{0}
            \subfloat[RPA Radial lower bounds]{\includegraphics[scale=0.6]{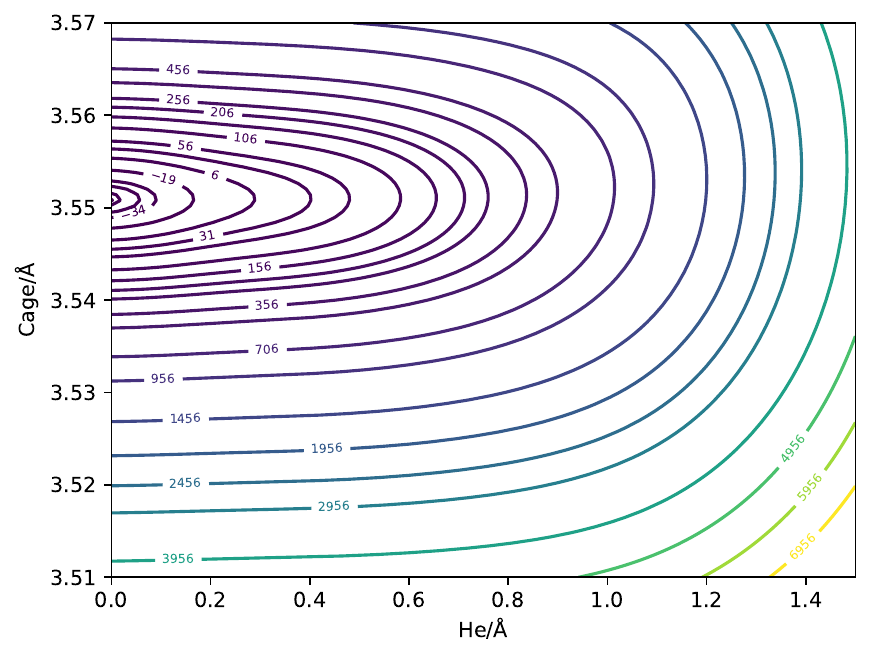}}\,
            \subfloat[RPA Angular lower bounds]{\includegraphics[scale=0.6]{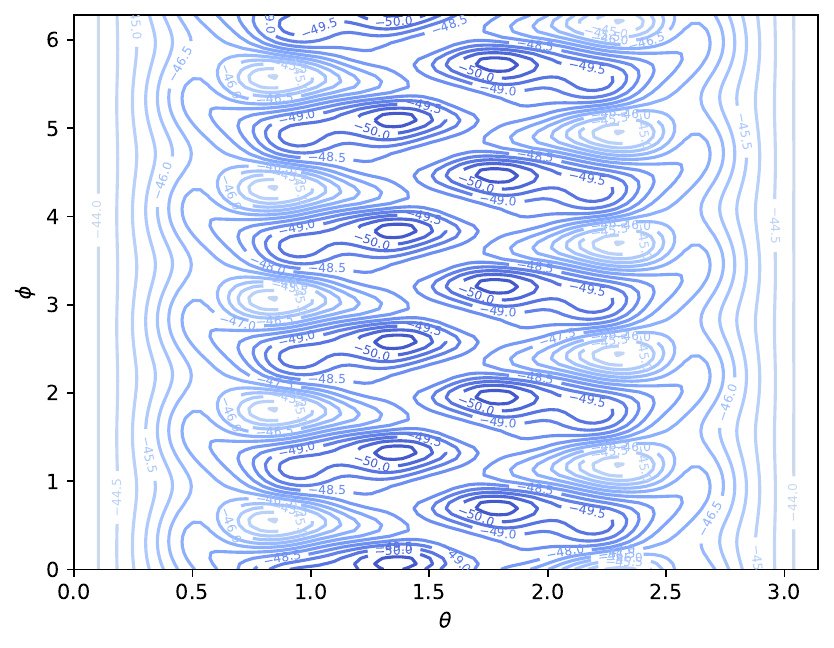}}\\
            \subfloat[RPA Radial upper bounds]{\includegraphics[scale=0.6]{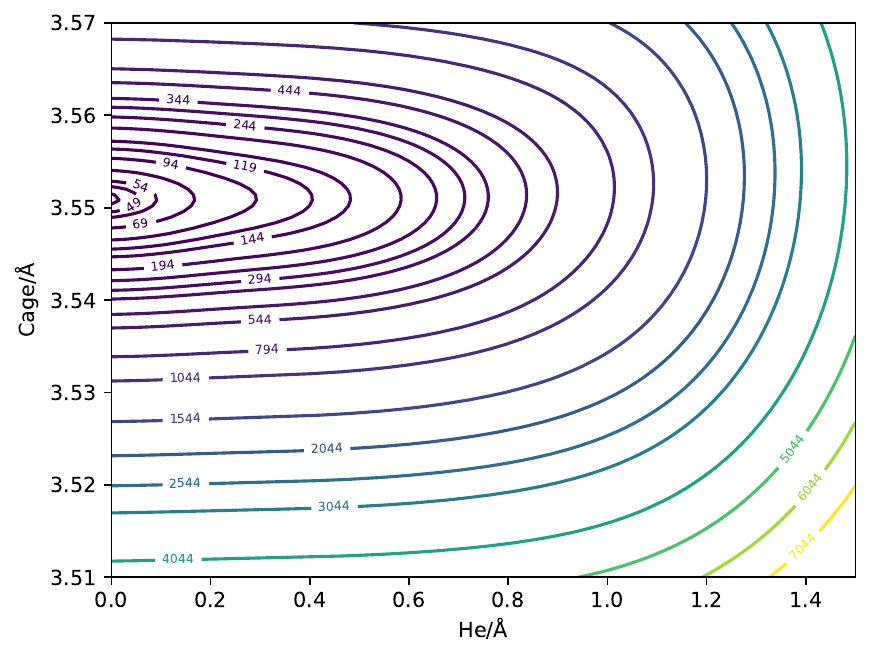}}\,
            \subfloat[RPA Angular upper bounds]{\includegraphics[scale=0.6]{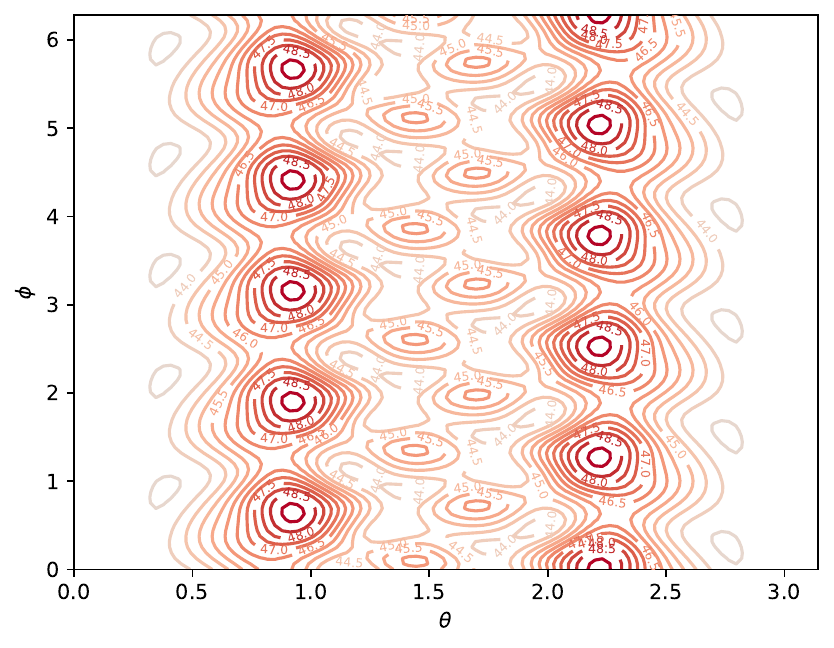}}
            \caption{95\% confidence interval on the value of the four-dimensional corrections to a fixed two-dimensional radial RPA@PBE PES: (a) radial lower bound, (b) angular lower bound, (c) radial upper bound, and (d) angular upper bound. The radial and angular slices are at the same fixed coordinates as in  Fig 2. Lower bound contours have their values shifted by -44\wavenumber, and upper bound by +44\wavenumber.
            }
            \label{fig: RPA confidence intervals}
        \end{figure*}

        \begin{table}
            \centering
            \subfloat[Bounds]{
                \begin{tabular}{cccccc}
                     \hline Kernel&Equation&$\sigma^2$&$r_{\ce{He}}$/\AA &$R_{\ce{C60}}$/\AA&Noise\\
                     \hline $K_{2D}$&(7)&(1e-5, 1e5)&(1e-5, 1e5)&(1e-5, 1e5)&(1e-9, 1e3)\\
                     $K_{4D}$&(8)&(1e-6, 1e-4)&(1e-1, 1e1)&(1e-1, 1e1)&(1e-8, 1e-4)\\
                     \hline
                \end{tabular}
            }\\
            \subfloat[$K_{2D}$]{
                \begin{tabular}{ccccc}
                    \hline Method &$\sigma^2$&$r_{\ce{He}}$/\AA &$R_{\ce{C60}}$/\AA&Noise\\
                    \hline MP2 &$4.07^2$&19.5&1.67&1e-9  \\
                    SCS-MP2 &$4.85^2$&20.9&1.82&1e-9  \\
                    SOS-MP2 &$5.27^2$&21.7&1.90&1e-9  \\
                    RPA@PBE &$5.44^2$&22.0&1.95&1e-9  \\
                    \hline
                \end{tabular}
            }\\
            \subfloat[$K_{4D}$]{
                \begin{tabular}{ccccc}
                    \hline Method &$\sigma^2$&$r_{\ce{He}}$/\AA &$R_{\ce{C60}}$/\AA&Noise\\
                    \hline All &$0.001^2$&10&10&1e-8  \\
                    \hline
                \end{tabular}
            }
            \caption{Hyperparameter (a) bounds and values for (b) 2D kernel and (c) 4D kernel used and found during the optimisation for each electronic structure method. Initial values are at the logarithmic centre of the bounds.}
            \label{table: Hyperparameters}
        \end{table}
        \begin{table}
            \centering
            \begin{tabular}{ccccc}
                 \hline Kernel & Metric& Angular PES & Lower bound & Upper bound  \\
                 \hline Mat\'{e}rn&Chordal&(-5.148, 6.718)&(-53.832, -41.556)&(43.072, 55.424)\\
                 Wendland-C2&Chordal&(-6.287, 15.505)&(-72.579, -39.925)&(44.133, 76.031)\\
                 Wendland-C2&Geodesic&(-5.153, 10.036)&(-57.579, -40.047)&(43.913, 60.431)\\
                 Wendland-C4&Chordal&(-6.056, 8.812)&(-59.632, -42.251)&(43.847, 61.952)\\
                 Wendland-C4&Geodesic&(-4.397, 5.012)&(-51.720, -42.127)&(42.716, 52.493)\\
                 \hline
            \end{tabular}
            \caption{Minimum and maximum values in \wavenumber of the MP2 angular PES and its lower and upper bound of the 95\% confidence interval for four-dimensional kernel for the varying choices of covariance function and metric.}
            \label{table: kernel metric comparison}
        \end{table}

        The set of 200 training points was generated by sampling each of the four coordinates from the following distributions
        \begin{align}
            r_{\ce{He}} &\sim N(0\text{\AA},  100\text{\wavenumber})\label{eq: train he}\\
            \theta &\sim U\left[0, \frac{\pi}{2}\right]\label{eq: train theta}\\
            \phi &\sim U\left[0, \frac{2\pi}{5}\right]\label{eq: train phi}\\
            R_{\ce{C60}} &\sim N(3.54\text{\AA},  500\text{\wavenumber}).\label{eq: train cage}
        \end{align}
        The angular distributions are chosen to force all the coordinates into a single spherical wedge. While this does not correspond to a uniform distribution on the surface of a sphere which may be preferable, this still generates a valid set of random training coordinates. Ensuring the angular coordinates lie within the given spherical wedge allows for ease of generating their symmetric equivalents.

        The distribution of the radial coordinates is motivated similarly to a potential optimised discrete variable representation (DVR) scheme.\cite{echavePotentialOptimizedDiscrete1992} For the means, the equilibrium position of \ce{He} within the cage, and cage radius of \ce{C60} are taken. For the variances, we use round values close to the harmonic frequency of each mode. These frequencies can be simply converted to squared distances modulated by the appropriate mass in the harmonic frequency calculation. A subtlety to note is that by definition $r_{\ce{He}}\geq0$ whereas sampling from the distribution in Eq \eqref{eq: train he} does not guarantee this. We choose to take the value as $|r_{\ce{He}}|$. This method generates sampling coordinates $r_{\ce{He}}\in (0, 1.5)$\AA, and $R_{\ce{C60}}\in(3.52, 3.57)$\AA. These ranges ensure the sampling points used when evaluating matrix elements in the DVR\cite{lightGeneralizedDiscreteVariable1985, lightDiscreteVariableRepresentationsTheir2000} lie within these intervals, allowing for sensible interpolation by the Gaussian Process (GP). 
        
        The two GPs were trained using a modified framework from scikit-learn.\cite{pedregosaScikitlearnMachineLearning2011} The training set was partitioned in a 2:1 ratio, with the larger set used to train a GP on the two-dimensional radial surface. The kernel given in Eq 7 used an anisotropic Mat\'{e}rn covariance function with $\nu=2.5$ with a white noise function added, with the hyperparameter bounds given in Table \ref{table: Hyperparameters}. Once the latter GP was trained, its latent surface was evaluated at the remaining training points, and an additional GP is trained on the four-dimensional surface representing the difference between the electronic structure calculation and the two-dimensional GP prediction. This smaller set of training points was symmetrised by applying the $I_h$ symmetry operations $C_5^n$ for $n\in [0, 1, 2, 3, 4]$ using the axis aligned along the $z$ direction as seen in Fig 1, and also the centre of inversion $i$, producing nine extra equivalent geometries with identical energies. The four-dimensional GP was thus trained on the symmetrised ``extended'' set of coordinates.
        
               The kernel given in Eq 8 used an anisotropic Mat\'{e}rn covariance function\cite{rasmussenGaussianProcessesMachine2005} with $\nu=2.5$ for $k_r$, but for $k_a$ given the multitude of choices for covariance function outlined in  Table I, with metrics in Eqns 9 and 10 the rationale for a specific choice is not obvious. All five choices (excluding Mat\'{e}rn function with geodesic metric due to its restriction on $\nu$) were tested by plotting the four-dimensional surfaces and the corresponding 95\% confidence interval for the MP2 data, analogously to Figs 2 and 3. The ranges of all these surfaces is given in Table \ref{table: kernel metric comparison} with the hyperparameter bounds and optimised values for this kernel also given in Table \ref{table: Hyperparameters}. These are noticeably tighter than their two-dimensional counterparts. This is due to the fact that the majority of the energy spread is accounted for in the two-dimensional GP and having wide bounds for the latter process often leads to ``ficitious'' LML minima (which are really very flat regions in the LML surface) to be chosen. Too large a length scale leads to a flat surface which ignores the locality of the data, whereas too low a length scale leads to overfitting. 

        Considering the range of the potential energy surface (PES), the Wendland-C2 covariance function performs worst, with a spread of over 15\wavenumber. Looking at its confidence interval, the lower and upper bound surfaces range over 30\wavenumber, which is not consistent with the other choices of covariance function and metric whose lower and upper bound surfaces of the confidence interval have the same range as the surface itself. Finally noticing the width of the confidence interval being well over 100\wavenumber allows us to discount the Wendland-C2 as a possible choice. The remaining three choices are all fairly similar to each other. Using the tightest 95\% confidence interval as the discriminator, this ranks Wendland-C4 with geodesic as best, closely followed by Mat\'{e}rn with chordal, and Wendland-C4 with chordal ranking last. Due to the high symmetry, the geodesic metric outperforming the chordal metric is due to its better ability at capturing the radial and angular patterns within the training data.\cite{gneitingStrictlyNonstrictlyPositive2013, padonouPolarGaussianProcesses2016}
        
        Analogously to the MP2 and RPA@PBE surfaces given in Fig 2, the equivalent SCS-MP2, SOS-MP2 and C(HF)-RPA PESs are shown in Fig \ref{fig: SCS SOS PES}. Looking at the radial surfaces, they once again are very smooth, indicative of being well described as a polynomial oscillator. There is appreciable anharmonicity along the $r_{\ce{He}}$ coordinate unlike the $R_{\ce{C60}}$ coordinate which does not show any. The alignment of the elliptical contours almost perfectly along each direction indicates there to be a lack of any observable coupling between the \ce{He} translational and \ce{C60} cage breathing mode. As with MP2 and RPA@PBE all these methods correctly predict the equilibrium \ce{He} position to be the origin. However, they also predict slightly different equilibrium cage radii for the \Endo{He}{C60} system, with SCS-MP2 predicting a cage radius of 3.539\AA\, SOS-MP2 one of 3.542\AA and C(HF)-RPA a radius of 3.536\AA.

        Turning our attention to the angular contours, they maintain the symmetry dictated by the $C_5$ axis as periodic $\frac{2\pi}{5}$ symmetry along the $\phi$ direction, and by the centre of inversion $i$ which manifests as an inversion about the point $(\theta, \phi) = (\frac{\pi}{2}, \pi)$. Once again there are regions of both positive (red) and negative (blue) corrections relative to the value at the origin. The locations of these regions is identical and consistent with the MP2 and RPA@PBE angular surfaces. The shape of the angular contours in all three MP2 methods is extremely similar. The only obvious difference between the SCS-MP2 and SOS-MP2 appears to be the former has a larger magnitude negative peak, whereas the latter has this in the positive contours. The range of these surfaces is larger than the RPA@PBE, but lower than the pure MP2. On the other hand, the C(HF)-RPA has a much smaller range, which is consistent with the RPA@PBE.
        
        Based on the information in Table \ref{table: kernel metric comparison}, we decided on Wendland-C4 covariance function with geodesic metric for the kernel due to it having the tightest 95\% confidence interval. The MP2 lower and upper bound surfaces are given in Fig 3, and the RPA@PBE equivalents are shown in Fig \ref{fig: RPA confidence intervals}. As with the MP2, both the radial lower and upper bounds have the same shape as each other, and the mean PES prediction. As the PES is agnostic to a linear energy shift, the behaviour of the GP is to strongly enforce the shape of the surface. The angular bounds on the other hand look very different to each other, and to their MP2 counterparts. While it maintains the appropriate symmetries, this once again points to the sensitivity of the four-dimensional kernel to its data. Both the lower and upper bound contours are centred on the value at the origin. This implies that the lower bound surface has its maximum at the origin, with the exact opposite for the upper-bound surface. However, the range of both the upper and lower bound surface is again on the order of 6\wavenumber, identical to the mean prediction which is also the case for the MP2 surface. This indicates the RPA@PBE surface has the tightest 95\% confidence interval.
            
    \section{Nuclear Hamiltonian Diagonalistion \label{secSI: Eigenstates}}
        \begin{figure*}
            \centering
            \setcounter{subfigure}{0}
            \subfloat[$s$]{\includegraphics[scale=0.23]{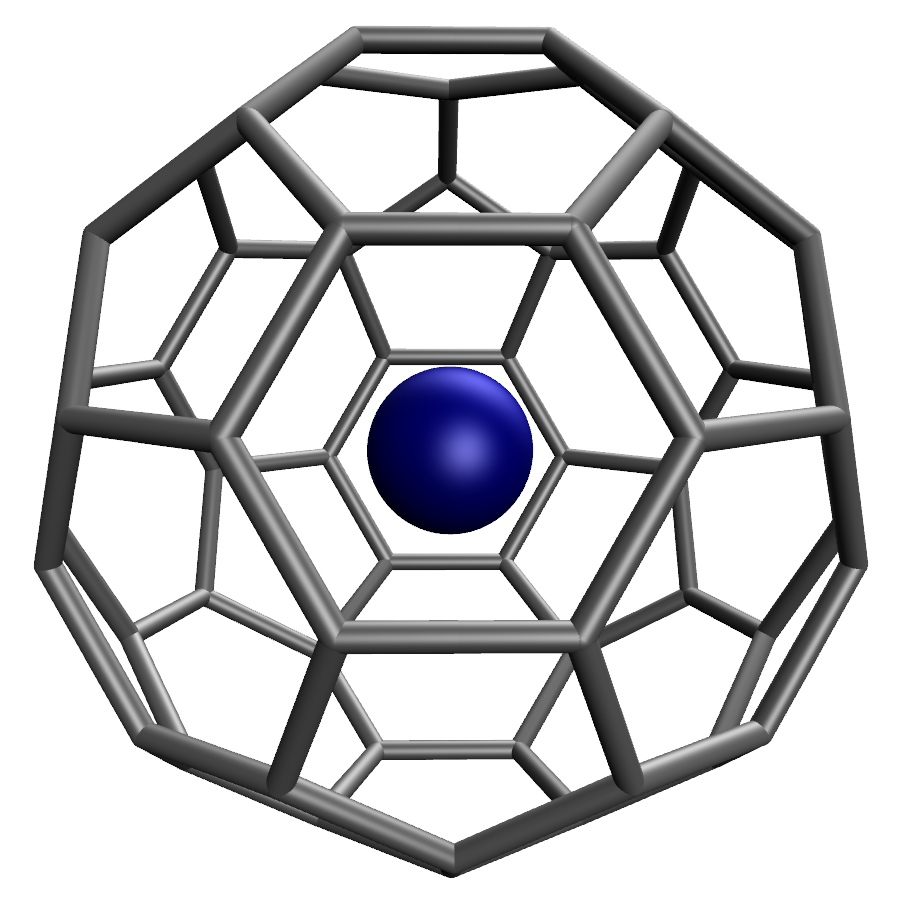}}\,
            \subfloat[$p$]{\includegraphics[scale=0.23]{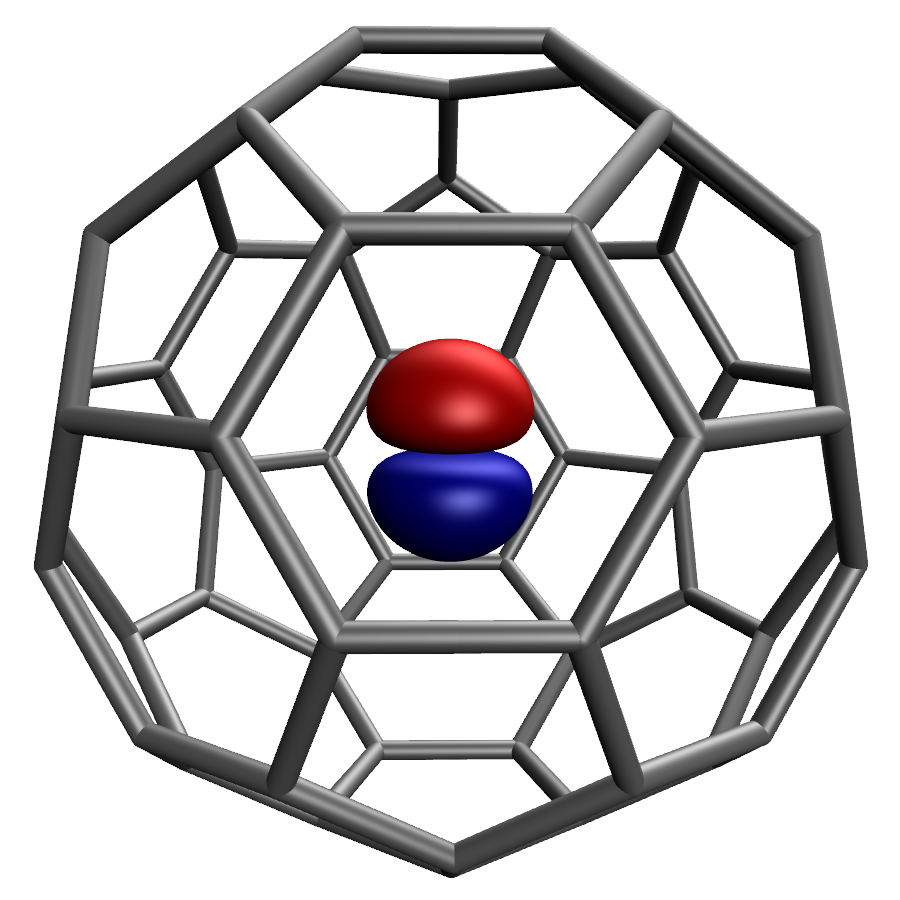}}\,
            \subfloat[$d$]{\includegraphics[scale=0.23]{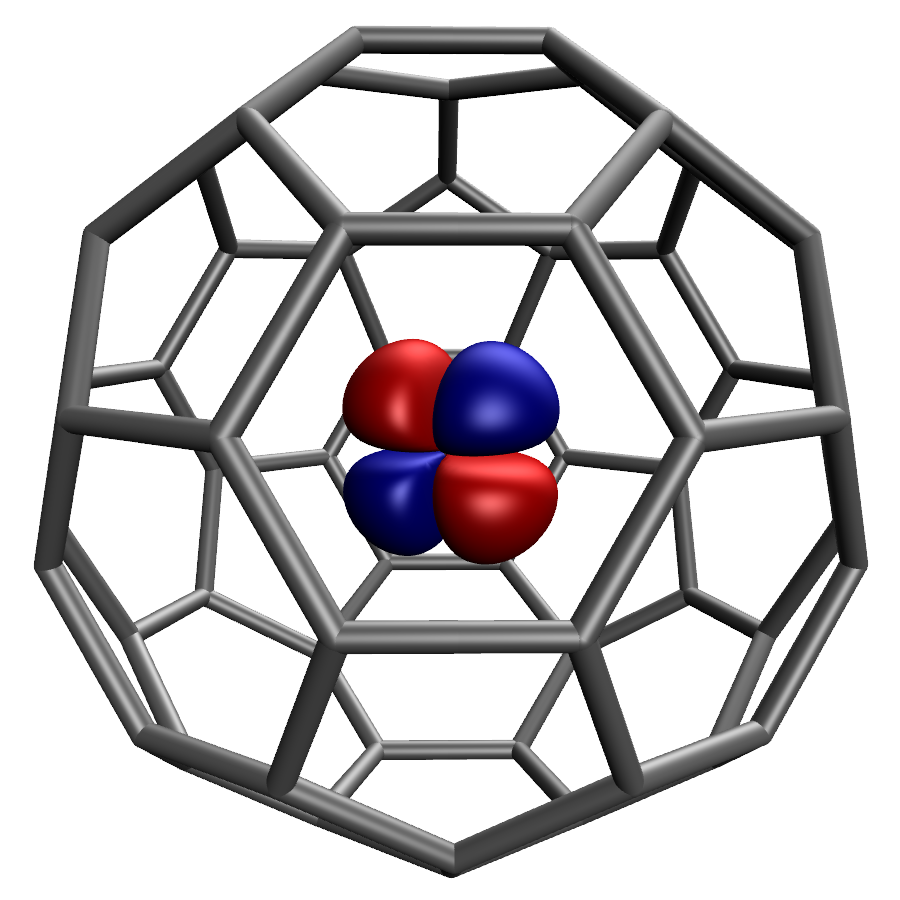}}\,
            \subfloat[$f$]{\includegraphics[scale=0.23]{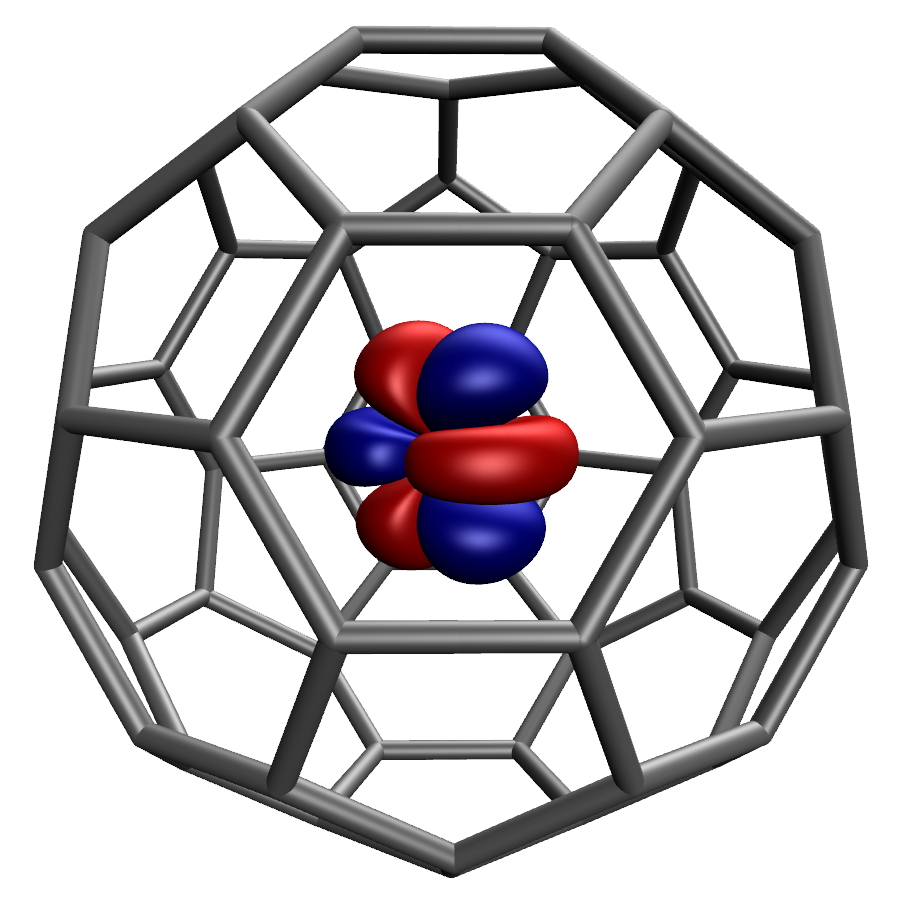}}
            \caption{RPA@PBE isosurfaces of (one of) the lowest lying set of (a) $s$, (b) $p$, (c) $d$, and (d) $f$ nuclear orbitals. The \ce{C60} cage has been rotated relative to Fig 1, with the $z$ direction pointing directly upwards, and $xy$ plane rotated to centre down a $C_3$ axis. The isosurfaces are taken at 0.5\% of the maximum amplitude of each wavefunction.}

            \label{fig: RPA Wavefunctions}
        \end{figure*}
        
        \begin{figure*}
        	\centering
        	\includegraphics[scale=1.2]{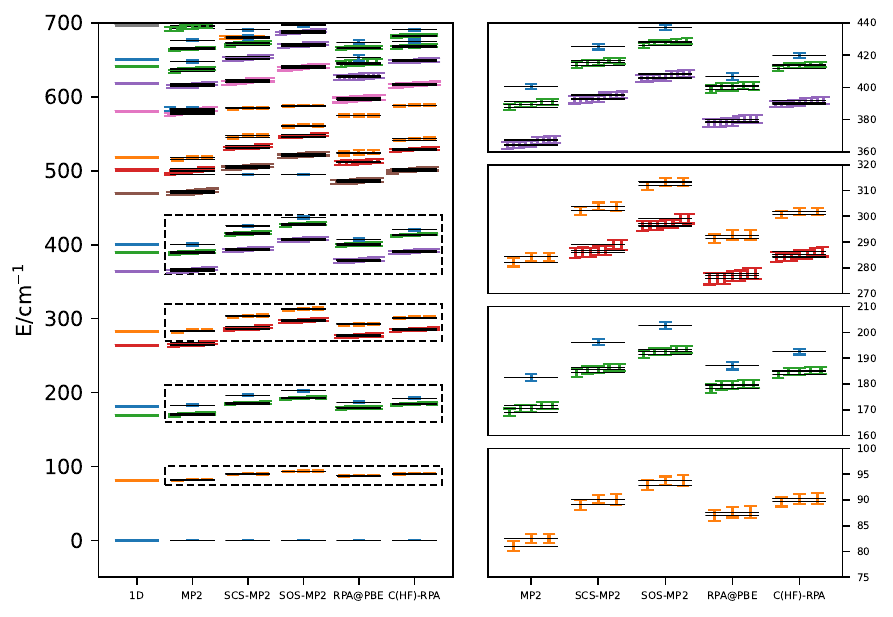}
        	\caption{Four-dimensional translational energies of \Endo{^4He}{C60} for MP2, SCS-MP2, SOS-MP2, RPA@PBE, and C(HF)-RPA PESs alongside the one-dimensional experimental derived counterparts. One-dimensional energies are coloured by angular momentum quantum number $l$ in increasing order: blue, orange, green, red, purple, brown, pink and grey. Error bars on the electronic structure methods are also coloured according to the near degeneracy of states.}
            \label{fig: He4 Translational Energies}
        \end{figure*}

        \begin{table}
            \centering
            \subfloat[$p$]{
                \begin{tabular}{c|ccccc}
                     $\bra{\downarrow}$, $\ket{\rightarrow}$& MP2 & SCS-MP2 & SOS-MP2 & RPA@PBE & C(HF)-RPA  \\
                     \hline MP2&-&0.04483&0.06520&0.02066&0.03726\\
                     SCS-MP2&0.04483&-&0.02044&0.03047&0.01000\\
                     SOS-MP2&0.06520&0.02044&-&0.05004&0.02933\\
                     RPA@PBE&0.02066&0.03047&0.05004&-&0.02171\\
                     C(HF)-RPA&0.03726&0.01000&0.02933&0.02171&-
                \end{tabular}
            }\\
            \subfloat[$d$]{
                \begin{tabular}{c|ccccc}
                     $\bra{\downarrow}$, $\ket{\rightarrow}$& MP2 & SCS-MP2 & SOS-MP2 & RPA@PBE & C(HF)-RPA  \\
                     \hline MP2&-&0.04697&0.06854&0.09546&0.10277\\
                     SCS-MP2&0.04697&-&0.02165&0.09157&0.08707\\
                     SOS-MP2&0.06854&0.02165&-&0.09740&0.08746\\
                     RPA@PBE&0.09546&0.09157&0.09740&-&0.02597\\
                      C(HF)-RPA&0.10277&0.08707&0.08746&0.02597&-
                \end{tabular}
            }\\
            \subfloat[$f$]{
                \begin{tabular}{c|ccccc}
                     $\bra{\downarrow}$, $\ket{\rightarrow}$& MP2 & SCS-MP2 & SOS-MP2 & RPA@PBE & C(HF)-RPA  \\
                     \hline MP2&-&0.09274&0.15994&0.10939&0.15879\\
                     SCS-MP2&0.09274&-&0.06847&0.19147&0.23331\\
                     SOS-MP2&0.15994&0.06847&-&0.25834&0.29830\\
                     RPA@PBE&0.10939&0.19147&0.25834&-&0.05513\\
                      C(HF)-RPA&0.15879&0.23331&0.29830&0.05513&-
                \end{tabular}
            }
            \caption{Hellinger distances of the (a) $p$, (b) $d$, and (c) $f$ state between all four electronic structure methods. Diagonal elements, by definition, are necessarily zero.}
            \label{table: Excited Hellinger}
        \end{table}

			Given the Hamiltonian and basis set defined in Eqns 11 and 14, generating the translational energies and wavefunctions for \Endo{He}{C60} involves calculating the matrix elements
        \begin{align}
            \braket{KLMB|\hat{H}|klmb}&=\braket{KLMB|\hat{h}^0|klmb}+\braket{KLMB|\Delta V|klmb} \nonumber\\
            &=\Braket{KLM|-\frac{1}{2m}\nabla^2_{\ce{He}}+\frac{1}{2}k_{\ce{He}}r^2|klm}\braket{B|b}\nonumber\\
            &+\Braket{B|-\frac{1}{2M}\nabla^2_{\ce{C60}}+\frac{1}{2}k_{\ce{C60}} (R_{\ce{C60}}-R_{\text{eq}})^2|b}\nonumber\\
            &\times \braket{KLM|klm}\nonumber\\
            &+\braket{KLMB|\Delta V|klmb}\nonumber\\
            &=\left[\left(2k+l+\frac{3}{2}\right)\omega_{\ce{He}}+\left(b+\frac{1}{2}\right)\omega_{\ce{C60}}\right]\times\nonumber\\
            &\delta_{Kk}\delta_{Ll}\delta_{Mm}\delta_{Bb}+\braket{KLMB|\Delta V|klmb} \label{eq: Hamiltonian matrix elements}
        \end{align}
		 where $\omega_{\ce{He}}$ and $\omega_{\ce{C60}}$ are the harmonic frequencies defined during the transformations given in Eqns 12 and 13, using a DVR\cite{lightGeneralizedDiscreteVariable1985, lightDiscreteVariableRepresentationsTheir2000} and potential optimised DVR\cite{echavePotentialOptimizedDiscrete1992} respectively. Due to the definitions of the basis functions, the $\hat{h}^0$ matrix is diagonal, whereas the $\Delta V$ matrix is not. However, as the basis functions are orthonormal, there exists an orthogonal transformation between the finite basis representation (FBR) and the DVR\cite{lightGeneralizedDiscreteVariable1985, lightDiscreteVariableRepresentationsTheir2000} which can be leveraged. The basis matrices
        \begin{align}
            T^{\ce{He}}_{(klm), (\alpha,\beta,\gamma)}&=\ket{kl}\ket{lm}=N_{klm}L_k^{l+\frac{1}{2}}(q^{\ce{He}}_{\alpha})Y_{lm}(\theta_{\beta}, \phi_{\gamma})\label{eq: He basis matrix}\\
            Y_{lm}(\theta, \phi)&=P_l^{m}(\cos(\theta))
                \begin{cases}
                    \cos(m\phi)\quad &m>0\\
                    1 \quad &m=0\\
                    \sin(|m|\phi)\quad& m<0
                \end{cases}
                \\
            T^{\ce{C60}}_{b, \delta}&=\ket{b}=N_bH_b(q^{\ce{C60}}_{\delta})\label{eq: C60 basis matrix}
        \end{align}

        constructed using generalised Laguerre polynomials $L_{k}^{l+\frac{1}{2}}$, real spherical harmonics $Y_{lm}$ built from generalised Laguerre polynomials $P_{l}^m$ and trignometric functions, and Hermite polynomials $H_b$. $N_{klm}$ and $N_b$ are the appropriate normalisation constants with the indices $(k,l,m,b)$ referring to the quantum numbers of the basis functions, and  $(\alpha, \beta, \gamma, \delta)$ refer to the Gaussian quadrature points. These also define the harmonic frequencies $\omega_{\ce{He}}$ and $\omega_{\ce{C60}}$. These can be combined to evaluate the potential matrix elements
        \begin{align}
            \mathbf{T}&=T^{\ce{He}}_{(klm), (\alpha,\beta,\gamma)} \otimes T^{\ce{C60}}_{b, \delta}\label{eq: Basis matrix}\\
            \braket{KLMB|\Delta V|klmb} &= \mathbf{T}\mathbf{V}^{DVR}\mathbf{T}^T\label{eq: DVR-FBR transformation}
        \end{align}

        where the $\mathbf{V}^{DVR}$ matrix is diagonal; entries being the potential sampled at the quadrature points minus the already accounted for harmonic contribution. Eq \eqref{eq: Basis matrix} leverages another advantage of DVR, as the overall quadrature grid is the outer product of sub-dimensional grids. This is not possible for Eq \eqref{eq: He basis matrix} due to the shared quantum number between the radial function and spherical harmonics, which in turns implies their quadrature grids are correlated. The $q^{\ce{He}}_{\alpha}$ grid is defined by generalised Gauss-Laguerre quadrature, $\cos(\theta_{\beta})$ and $\phi_{\gamma}$ are defined on Gauss-Laguerre grids, and $q^{\ce{C60}}_{\delta}$ is defined by a Gauss-Hermite grid. Calculating this matrix product is equivalent to evaluating the integral using Gaussian quadrature.\cite{lightGeneralizedDiscreteVariable1985,lightDiscreteVariableRepresentationsTheir2000} Usually, the number of quadrature points and basis functions are equal, making the basis matrices in Eqns \eqref{eq: He basis matrix} and \eqref{eq: C60 basis matrix} square, but this does not need to be the case. Due to coupling of radial functions with differing $l$ values, and oscillatory nature of the spherical harmonics, in order to converge the integrals, twice the number of radial functions are used as quadrature points in the  $q^{\ce{He}}_{\alpha}$ grid, and four times the number of angular functions are used as quadrature points in the $\phi_{\gamma}$ grid. This factor of eight between the number of basis functions and quadrature points implies working in the variational basis representation (VBR)\cite{lightGeneralizedDiscreteVariable1985, lightDiscreteVariableRepresentationsTheir2000}, as opposed to the FBR, as more matrix elements are converged. 
        
        While the energies for all four electronic structure methods are given in Fig 4, only the ground state wavefunctions are compared in Tables II and III. For MP2, one of the lowest lying set of $s$, $p$, $d$, and $f$ orbitals is given in Fig 5. The RPA@PBE analogue is given here in Fig \ref{fig: RPA Wavefunctions}. Once again, they have strikingly regular patterns, with assignment of translational and angular momentum quantum numbers $k$ and $l$ respectively being very simple by analysing the nodal patterns. With the cage being reoriented, neither the $d$ nor $f$ orbital lie perfectly along the $x$ and $y$ axes. The ordering of the nearly degenerate orbitals is subtly different between the RPA@PBE and MP2, which is due to the difference in angular contours seen in Fig \ref{fig: PES}.
        
        The Hellinger distances of the $p$, $d$ and $f$ states of the MP2 in Fig 5 and RPA@PBE in Fig \ref{fig: RPA Wavefunctions} are compared to the SCS-MP2, SOS-MP2 and C(HF)-RPA in Table \ref{table: Excited Hellinger}. Similar to the ground $s$ state in Table III, the MP2 and RPA@PBE $p$ states are very similar to each other, as are the SCS-MP2 and SOS-MP2. Once again, the C(HF)-RPA is very close with the SCS-MP2. However, this is not found to be the case in the $d$ state distances. Despite the energy ordering of MP2$<$RPA@PBE$<$C(HF)-RPA$<$SCS-MP2$<$SOS-MP2, all the MP2 wavefunctions are closer together than the RPA methods. This is justifiable considering the higher fold degeneracy of the $d$ basis states, which are the major contributor to the eigenfunction and their exact contribution being sensitive to the shape of the PES. This is further exacerbated within the $f$ nuclear orbital distances, as the energy matches are poorer but the wavefunction distances are even further apart. This can be indicative of nuclear orbitals for each method having different principal $x$ and $y$ axes. If they have been rotated a different amount relative to each other, this will lead to a poorer overlap of the orbitals, leading to a larger Hellinger distance.
			While the results in Section III are for \Endo{^3He}{C60}, the translational eigenstate energies for \Endo{^4He}{C60} are shown in Fig \ref{fig: He4 Translational Energies}, which is the analogous figure to Fig \ref{fig: Translational Energies}. As for the lighter isotope, C(HF)-RPA and the semi-empirical SCS-MP2 and SOS-MP2 methods consistently overestimate the energies; MP2 and RPA@PBE perform significantly better. Once again, the MP2 reaches spectroscopic accuracy with the one-dimensional results, with the RPA@PBE lagging a few wavenumber behind.

    \bibliography{HeC604D}